\documentclass[aps,prd,12pt,nofootinbib,superscriptaddress,longbibliography]{revtex4-2}
\usepackage{bm,amsmath,amssymb,graphicx,array,booktabs}
\usepackage{xcolor,hyperref,CJK,orcidlink}
\hypersetup{colorlinks=true,linkcolor=blue,citecolor=blue,urlcolor=blue}
\usepackage{ulem}

\begin{document}

\begin{CJK*}{UTF8}{gbsn}

\title{Analytic Electromagnetic Effective Source for Geodesic Orbit in Schwarzschild Spacetime}
\author{Xuchen Lu (路旭晨) 
\orcidlink{0000-0002-9093-9059}}
\email{luxuchen@nbu.edu.cn}
\affiliation{Zhejiang Key Laboratory of Extreme Universe,
Institute of Fundamental Physics and Quantum Technology,
Department of Physics, School of Physical Science and Technology, Ningbo
University, Ningbo, Zhejiang 315211, China}

\author{Yungui Gong (龚云贵)
\orcidlink{0000-0001-5065-2259}}
\email{gongyungui@nbu.edu.cn}
\affiliation{Zhejiang Key Laboratory of Extreme Universe,
Institute of Fundamental Physics and Quantum Technology, Department of Physics, School of Physical Science and Technology, Ningbo University, Ningbo, Zhejiang 315211, China}

\author{Bokai Zhang (张博凯)
\orcidlink{0009-0004-8123-1894}}
\email{bkzhang07@163.com}
\affiliation{School of Physics, Huazhong University of Science and Technology, 1037 LuoYu Rd, Wuhan, Hubei 430074, China}

\author{Chao Zhang (张超)
\orcidlink{0000-0001-8829-1591}}
\email{zhangchao1@nbu.edu.cn}
\affiliation{Zhejiang Key Laboratory of Extreme Universe,
Institute of Fundamental Physics and Quantum Technology, Department of Physics, School of Physical Science and Technology, Ningbo University, Ningbo, Zhejiang 315211, China}

\author{Wenting Zhou (周文婷)
\orcidlink{0000-0003-4046-753X}}
\email{zhouwenting@nbu.edu.cn}
\affiliation{Zhejiang Key Laboratory of Extreme Universe,
Institute of Fundamental Physics and Quantum Technology, Department of Physics, School of Physical Science and Technology, Ningbo University, Ningbo, Zhejiang 315211, China}

\begin{abstract}
Self-force effects are essential for accurately describing the motion of small bodies in curved spacetime.
The effective-source approach privides a promising route to higher-order self-force calculations. 
In this paper, we construct analytic electromagnetic effective source for geodesic motion in Schwarzschild spacetime.
An angular completion and a tetrad decomposition express the vector-harmonic puncture modes as finite sums of scalar modes.
We derive the coupling coefficients for both parity sectors and obtain the effective sources from the Maxwell mode equations.
The point-particle delta contribution cancels, and representative eccentric-orbit results show finite mode sources near the particle.
These expressions provide the source terms for future $1+1$-dimensional residual-field calculations.
\end{abstract}
\maketitle
\end{CJK*}

\section{Introduction}

The interaction of a charged particle with its own electromagnetic field
is a fundamental problem in classical electrodynamics.
The field generated by the particle carries energy and momentum and,
in turn, modifies its motion.
In the point-particle description, however, the electromagnetic field
diverges on the worldline, making a direct evaluation of the
self-interaction ill-defined.
A consistent treatment must therefore separate the singular contribution
from the finite field responsible for the self-force.
In flat spacetime, Dirac developed a covariant formulation of this
problem by analyzing the retarded and advanced fields and absorbing the
divergent contribution into the particle's mass
\cite{Dirac:1938nz}.
The resulting Abraham--Lorentz--Dirac equation established a relativistic
description of electromagnetic radiation reaction and provided a
foundation for subsequent investigations in curved spacetime.

The extension to curved spacetime was developed by DeWitt and Brehme
\cite{DeWitt:1960fc}, with a correction to the local curvature
contribution subsequently supplied by Hobbs \cite{hobbs1968vierbein}.
A central feature of the curved-spacetime problem is that the retarded
Green function generally has support inside the light cone.
Consequently, the field acting on the particle contains a tail
contribution that depends on its past motion.
The electromagnetic self-force is therefore generally nonlocal in time,
and its evaluation requires information about field propagation through
the background geometry.
Quinn and Wald later obtained the electromagnetic equation of motion
from an axiomatic approach \cite{Quinn:1996am}.
These developments established the theoretical basis of the self-force
problem, while also revealing the difficulty of translating its formal
solution into practical calculations.

A useful description of the self-interaction is provided by the
Detweiler--Whiting decomposition \cite{Detweiler:2002mi}.
In this formulation, the retarded field is separated into a singular
field, which reproduces the local divergence, and a regular field,
which satisfies the homogeneous field equation and determines the
self-force.
The regular field accounts for both dissipative and conservative
effects, so the self-force contains information beyond that obtained
from radiative fluxes alone.
For particles moving in curved spacetimes, the central computational
task is thus to extract this regular field from a retarded solution
sourced by a point particle.

Several regularization schemes have been developed to carry out this
extraction.
In the mode-sum method,
analytically determined singular contributions are subtracted from the spherical-harmonic modes before the
regularized sum is evaluated \cite{Barack:1999wf,Barack:2001bw}.
This method has been applied to electromagnetic self-force calculations for particles on eccentric geodesics in Schwarzschild spacetime \cite{Haas:2011np} and 
equatorial circular orbit in Kerr spacetime \cite{Torres:2020fye}.
Mode-sum regularization has also been used in first-order scalar
self-force calculations \cite{Detweiler:2002gi} and in gravitational calculations for circular and eccentric Schwarzschild orbits and generic Kerr orbits
\cite{Barack:2007tm,Detweiler:2008ft,Akcay:2010dx,Barack:2010tm,Akcay:2013wfa,Osburn:2014hoa,vandeMeent:2017bcc}.
These results have supported models of self-forced inspiral motion
\cite{Pound:2009sm,Warburton:2011fk,Osburn:2015duj,Warburton:2017sxk,VanDeMeent:2018cgn,Lynch:2021ogr,Lynch:2023gpu,Zhang:2026zqs}.
The precision sought for long inspirals has further motivated second-order self-force calculations
\cite{Rosenthal:2006nh,Rosenthal:2006iy,Pound:2012nt,Gralla:2012db,Pound:2017psq,Pound:2019lzj,Miller:2020bft,Wardell:2021fyy,Upton:2021oxf,Upton:2023tcv}.
However,
the standard mode-sum method relies on the retarded field's $1/r$ singularity, whose individual angular modes remain finite at the particle.
At second order, the field is generally more singular near the particle than at first order. 
Individual angular modes can then diverge at the worldline, so the standard first-order mode-sum prescription cannot be applied directly \cite{Pound:2014xva,Wardell:2015ada}.
An alternative is to approximate the singular field locally by a puncture and subtract it from the retarded field.
The residual field then satisfies an equation in which the leading singular contributions cancel.
This is the effective-source method \cite{Barack:2007jh,Vega:2007mc}.

Early effective-source calculations in Schwarzschild spacetime focused mainly on circular orbits, for both scalar fields and Lorenz-gauge gravitational perturbations \cite{Vega:2007mc,Warburton:2013lea,Wardell:2015ada}.
The method has also been used for the self-consistent evolution of a scalar charge and its orbit \cite{Diener:2011cc}.
Frequency-domain methods are well suited to bound motion, but extending effective-source calculations from circular to eccentric orbits presents two distinct challenges.
First, the puncture and effective-source modes must be constructed for all orbital phases. For a circular orbit, the particle radius is fixed and the radial velocity vanishes.
For an eccentric orbit, the radius varies and the radial velocity reverses sign at the turning points.
Radial-velocity terms enter the local puncture and its angular projection, while the mode operators differentiate orbit-dependent quantities.
Deriving the eccentric-orbit effective source is therefore substantially more involved than in the circular case.
The second challenge arises in the frequency domain.
At a fixed radius between periastron and apastron, the particle passes twice per radial period, causing the physical puncture and effective source to switch between interior and exterior branches.
Their limited differentiability at these crossings leads to algebraically decaying radial Fourier coefficients.
Accurately reconstructing the source and its derivatives near the particle may therefore require many harmonics.
An eccentric-orbit self-force calculation must consequently control both the accuracy of the effective source and the error from truncating the radial-harmonic sum.

Leather and Warburton addressed this convergence problem for a scalar charge by analytically extending the two source branches before Fourier decomposition, developing the extended effective-source (EES) method \cite{Leather:2023dzj}.
For gravitational perturbations, Zhang et al. constructed analytic effective sources for generic Schwarzschild geodesics by using an angular completion and a tetrad decomposition to express tensor-harmonic puncture modes as finite sums of scalar modes \cite{Zhang:2025eqz}.
Their construction was subsequently used in a time-domain gravitational self-force calculation for circular orbits \cite{Zhang:2026gdk}.
Lu et al. then extended the EES method to the coupled Lorenz-gauge gravitational equations.
Using the generic-orbit sources of Zhang et al.,
they computed the gravitational self-force for eccentric Schwarzschild orbits in the frequency domain \cite{Lu:2026jzh}.

Despite these advances in the gravitational case, the electromagnetic effective source requires a separate construction.
The electromagnetic puncture is linear, rather than quadratic, in the parallel-transported four-velocity, and its vector-harmonic projection has different normalization and parity couplings.
Thus, the electromagnetic effective source cannot be obtained from the gravitational one by simply reducing its tensor rank.
In this paper, we construct analytic effective-source for generic Schwarzschild geodesics.
An angular completion and a complex tetrad reduce the vector-harmonic puncture modes to finite sums of scalar modes.
We derive the coupling coefficients and apply the Maxwell mode operators to obtain the effective sources.
These results provide the analytic foundation for future electromagnetic self-force calculations using the EES method.

The paper is organized as follows.
Section II introduces the Maxwell mode equations, the effective-source formulation, and the local electromagnetic puncture.
Section III develops the tetrad decomposition and derives the analytic relations between scalar and vector-harmonic puncture modes.
Section IV constructs the effective-source modes and presents representative numerical examples.
Section V concludes.
Throughout, we use geometrized units $G=c=1$, set the black-hole mass $M=1$, and adopt the metric signature $(-,+,+,+)$.

\section{Electromagnetic effective-source formalism}
\label{sec:Ele_es}

\subsection{Maxwell equations on a Schwarzschild background}
\label{sec:Maxwell_eqs}

In Schwarzschild coordinates $x^\mu=(t,r,\theta,\phi)$, the line element is
\begin{equation}
ds^2=-fdt^2+f^{-1}dr^2+r^2\left(d\theta^2+\sin^2\theta\,d\phi^2\right),
\label{eq:schwarzschild-metric}
\end{equation}
where $f=1-2/r$.
The inhomogeneous Maxwell equation in the curved spacetime is
\begin{equation}
\label{eq:maxwell-equation}
\nabla_\nu F^{\mu\nu}=4\pi j^\mu,
\end{equation}
where $F_{\mu\nu}$ is the field-strength tensor, 
\begin{equation}
\label{eq:field-strength}
F_{\mu\nu}
=\nabla_\mu A_\nu-\nabla_\nu A_\mu,
\end{equation}
and 
$A_\mu$ is the electromagnetic vector potential.
The point-particle current is
\begin{equation}
\label{eq:point-particle-current}
j^\mu(x)=q\int_{-\infty}^{\infty}u^\mu(\tau)\frac{\delta^{(4)}[x-x_p(\tau)]}{\sqrt{-g}} d\tau.
\end{equation}
Here the subscript $p$ labels the particle's worldline,
$g=-r^4\sin^2\theta$ is the metric determinant,
and the current satisfies $\nabla_\mu j^\mu=0$ in the distributional sense.
The worldline $x_p^\mu(\tau)$ is parametrized by proper time $\tau$, with four-velocity $u^\mu=dx_p^\mu/d\tau$ normalized by $u^\mu u_\mu=-1$.
On a constant-$t$ hypersurface, the current becomes
\begin{equation}
\label{eq:point-particle-current-coordinate}
j^\mu(t,r,\theta,\phi)
=\frac{q u^\mu}{u^t r^2\sin\theta}\delta[r-r_p(t)]\delta[\theta-\theta_p(t)]\delta[\phi-\phi_p(t)].
\end{equation}

Substituting Eq. \eqref{eq:field-strength} into Eq. \eqref{eq:maxwell-equation} and imposing the Lorenz gauge condition $\nabla_\mu A^\mu=0$ gives
\begin{equation}
\Box A_\mu-R_\mu{}^\nu A_\nu=-4\pi j_\mu,
\label{eq:maxwell-wave-equation-general}
\end{equation}
where $\Box\equiv\nabla_\alpha\nabla^\alpha$.
Since the Schwarzschild spacetime is Ricci flat, $R_{\mu\nu}=0$, this reduces to
\begin{equation}
\label{eq:maxwell-vector-potential}
\Box A_\mu=-4\pi j_\mu.
\end{equation}
We introduce the covariant vector spherical harmonics \cite{Rosa:2011my}
\begin{equation}
\begin{split}
Z_\mu^{(1)\ell m}
={}&
\left(1,0,0,0\right)Y_{\ell m},
\\
Z_\mu^{(2)\ell m}
={}&
\left(0,f^{-1},0,0\right)Y_{\ell m},
\\
Z_\mu^{(3)\ell m}
={}&
\frac{r}{\sqrt{\lambda}}
\left(0,0,\partial_\theta,\partial_\phi\right)Y_{\ell m},
\\
Z_\mu^{(4)\ell m}
={}&
\frac{r}{\sqrt{\lambda}}
\left(
0,0,
\frac{1}{\sin\theta}\partial_\phi,
-\sin\theta\,\partial_\theta
\right)Y_{\ell m},
\end{split}
\label{eq:vector-harmonics}
\end{equation}
where $\lambda\equiv\ell(\ell+1)$.
For $\ell=0$, only $Z_\mu^{(1)00}$ and $Z_\mu^{(2)00}$ are nonzero.
These harmonics are orthonormal with respect to the auxiliary positive-definite tensor $\eta^{\mu\nu}$,
\begin{equation}
\int
\left(Z_\mu^{(i)\ell m}\right)^*
\eta^{\mu\nu}
Z_\nu^{(j)\ell' m'}
\,d\Omega=
\delta^{ij}\delta_{\ell\ell'}\delta_{mm'},
\label{eq:vector-harmonic-orthogonality}
\end{equation}
where $d\Omega=\sin\theta\,d\theta\,d\phi$ and
$\eta^{\mu\nu}=\operatorname{diag}\left(1,f^2,r^{-2},r^{-2}\sin^{-2}\theta\right)$.
The tensor $\eta^{\mu\nu}$ defines the harmonic inner product and is not the inverse spacetime metric $g^{\mu\nu}$.
Under the parity transformation
$(\theta,\phi)\mapsto(\pi-\theta,\phi+\pi)$,
the harmonics $Z_\mu^{(1)\ell m}$,
$Z_\mu^{(2)\ell m}$, and
$Z_\mu^{(3)\ell m}$ have parity $(-1)^\ell$ and form the even-parity sector.
The harmonic $Z_\mu^{(4)\ell m}$ has parity $(-1)^{\ell+1}$ and forms the odd-parity sector.

The vector potential and covariant current are expanded as
\begin{equation}
\label{eq:Amu}
A_\mu(t,r,\theta,\phi)=\frac{1}{r}\sum_{\ell=0}^{\infty}\sum_{m=-\ell}^{\ell}\sum_{i\in I_\ell}c_i\mathcal A_{\ell m}^{(i)}(t,r)Z_\mu^{(i)\ell m}(\theta,\phi),
\end{equation}
\begin{equation}
j_\mu(t,r,\theta,\phi)=\frac{1}{r}\sum_{\ell=0}^{\infty}\sum_{m=-\ell}^{\ell}\sum_{i\in I_\ell}c_i\mathcal J_{\ell m}^{(i)}(t,r)Z_\mu^{(i)\ell m}(\theta,\phi),
\end{equation}
where
\begin{equation}
I_0
=\{1,2\},\quad
I_{\ell\geq1}
=\{1,2,3,4\},\quad
c_1=c_2=1,\quad
c_3=c_4=\lambda^{-1/2}.
\end{equation}
The constants $c_i$ fix the normalization of the mode amplitudes.
Equation \eqref{eq:vector-harmonic-orthogonality} gives
\begin{equation}
\mathcal J_{\ell m}^{(i)}(t,r)
=
\frac{r}{c_i}
\int
\left(Z_\mu^{(i)\ell m}\right)^*
\eta^{\mu\nu}j_\nu
\,d\Omega.
\label{eq:current-mode-projection}
\end{equation}
For the point-particle current, we define the source coefficients by
\begin{equation}
-4\pi f\mathcal J_{\ell m}^{(i)}(t,r)
=\mathcal S_{\ell m}^{(i)}(t)\delta[r-r_p(t)].
\end{equation}

For convenience,
we introduce the tortoise coordinate
\begin{equation}
\label{eq:tortoise-coordinate}
\frac{dr_*}{dr}=f^{-1},\quad
r_*=r+2\ln\left(\frac{r}{2}-1\right).
\end{equation}
Substituting Eq. \eqref{eq:Amu} into Eq. \eqref{eq:maxwell-vector-potential} and projecting onto the vector-harmonic basis gives
\begin{equation}
\begin{split}
\mathcal D_\ell\mathcal A_{\ell m}^{(1)}
+\frac{2}{r^2}
\left(
\partial_t\mathcal A_{\ell m}^{(2)}
-\partial_{r_*}\mathcal A_{\ell m}^{(1)}
\right)
={}&
\mathcal S_{\ell m}^{(1)}
\delta[r-r_p(t)],
\\
\mathcal D_\ell\mathcal A_{\ell m}^{(2)}
+\frac{2}{r^2}
\left[
\partial_t\mathcal A_{\ell m}^{(1)}
-\partial_{r_*}\mathcal A_{\ell m}^{(2)}
-f^2\left(
\mathcal A_{\ell m}^{(2)}
-\mathcal A_{\ell m}^{(3)}
\right)
\right]
={}&
\mathcal S_{\ell m}^{(2)}
\delta[r-r_p(t)],
\\
\mathcal D_\ell\mathcal A_{\ell m}^{(3)}
+\frac{2f\lambda}{r^2}
\mathcal A_{\ell m}^{(2)}
={}&
\mathcal S_{\ell m}^{(3)}
\delta[r-r_p(t)],
\\
\mathcal D_\ell\mathcal A_{\ell m}^{(4)}
={}&
\mathcal S_{\ell m}^{(4)}
\delta[r-r_p(t)].
\end{split}
\label{eq:maxwell-mode-equations}
\end{equation}
where $\mathcal D_\ell\equiv-\partial_t^2+\partial_{r_*}^2-f\frac{\lambda}{r^2}$.
The components $i=1,2,3$ form the coupled even-parity sector, while the odd-parity component $i=4$ obeys a decoupled equation.
For $\ell=0$, only the equations for $\mathcal A_{00}^{(1)}$ and $\mathcal A_{00}^{(2)}$ remain.
The Lorenz gauge condition becomes
\begin{equation}
-\partial_t\mathcal A_{\ell m}^{(1)}
+\partial_{r_*}\mathcal A_{\ell m}^{(2)}
+\frac{f}{r}
\left(
\mathcal A_{\ell m}^{(2)}
-\mathcal A_{\ell m}^{(3)}
\right)
=0.
\label{eq:lorenz-mode-condition}
\end{equation}
This constraint involves only the even-parity amplitudes. 

The reality of the vector potential implies
\begin{equation}
\mathcal A_{\ell,-m}^{(i)}
=
(-1)^m
\left(
\mathcal A_{\ell m}^{(i)}
\right)^*,
\label{eq:mode-reality-condition}
\end{equation}
with analogous relations for the current and source modes.

\subsection{Effective-source approach}
\label{subsec:effective-source-approach}

The electromagnetic self-force in curved spacetime was derived by DeWitt and Brehme \cite{DeWitt:1960fc} and corrected by Hobbs \cite{hobbs1968vierbein}.
The Detweiler--Whiting decomposition \cite{Detweiler:2002mi} separates the retarded vector potential into singular and regular parts,
\begin{equation}
A_\mu^{\mathrm{ret}}
=
A_\mu^{\mathrm R}
+
A_\mu^{\mathrm S},
\label{eq:em-DW-decomposition}
\end{equation}
where $A_\mu^{\mathrm S}$ contains the local singular structure and
the smooth regular field $A_\mu^{\mathrm R}$ determines the
electromagnetic self-force.
In the Lorenz gauge on the Schwarzschild background,
the retarded and singular fields satisfy
\begin{equation}
\Box A_\mu^{\mathrm{ret}}
=
-4\pi j_\mu,
\qquad
\Box A_\mu^{\mathrm S}
=
-4\pi j_\mu,
\label{eq:em-retarded-singular-equations}
\end{equation}
while the regular field satisfies
\begin{equation}
\Box A_\mu^{\mathrm R}=0.
\label{eq:em-regular-equation}
\end{equation}

The singular field is known locally through an expansion about the
particle worldline.
In the effective-source approach, it is replaced by a puncture
$A_\mu^{\mathcal P}$ that reproduces the singular field to a prescribed
order \cite{Heffernan:2012su}.
We define the residual vector potential by
\begin{equation}
A_\mu^{\mathcal R}
\equiv
A_\mu^{\mathrm{ret}}
-
A_\mu^{\mathcal P}.
\label{eq:em-residual-definition}
\end{equation}
The superscripts $\mathrm R$ and $\mathcal R$ distinguish the exact
Detweiler--Whiting regular field from the residual field associated
with the finite-order puncture.
Their difference is
\begin{equation}
A_\mu^{\mathcal R}-A_\mu^{\mathrm R}
=
A_\mu^{\mathrm S}-A_\mu^{\mathcal P},
\label{eq:regular-residual-difference}
\end{equation}
and is controlled by the puncture order.

Substituting Eq. \eqref{eq:em-residual-definition} into
Eq. \eqref{eq:maxwell-vector-potential} gives
\begin{equation}
\Box A_\mu^{\mathcal R}
=
-4\pi j_\mu
-
\Box A_\mu^{\mathcal P}
\equiv
S_\mu^{\mathrm{eff}}.
\label{eq:em-effective-source-schwarzschild}
\end{equation}
The distributional contribution from the point-particle current is
cancelled by the corresponding contribution from the puncture.
The regularity of the remaining effective source is determined by
the puncture order.

Although the reconstructed retarded field is in the Lorenz gauge,
the finite-order puncture need not satisfy the Lorenz condition
separately.
The residual field therefore obeys
\begin{equation}
\nabla^\mu A_\mu^{\mathcal R}
=
-\nabla^\mu A_\mu^{\mathcal P}.
\label{eq:residual-lorenz-constraint}
\end{equation}
Imposing a homogeneous Lorenz condition on the residual field alone would generally be inconsistent with the decomposition in Eq. \eqref{eq:em-residual-definition}.

\subsection{Local electromagnetic puncture}

We now construct the local electromagnetic puncture.
For an equatorial orbit, let
\begin{equation}
z_p^\mu
=
(t_p,r_p,\pi/2,\phi_p)
\label{eq:particle-reference-point}
\end{equation}
be the reference point on the worldline.
On the equal-time hypersurface $t=t_p$, we define
\begin{equation}
\Delta x^\mu
\equiv
x^\mu-z_p^\mu
=
\left(
0,\Delta r,\Delta\theta,\Delta\phi
\right),
\label{eq:coordinate-separation}
\end{equation}
where
$\Delta r=r-r_p$,
$\Delta\theta=\theta-\pi/2$,
and
$\Delta\phi=\phi-\phi_p$.
The azimuthal difference is taken on the local branch that vanishes
at the particle.
We use $\rho$ to count powers of the coordinate separation, with
$\Delta x^\mu=O(\rho)$.

For geodesic motion, the Lorenz-gauge Detweiler--Whiting singular
vector potential has the local expansion \cite{Heffernan:2012su}
\begin{equation}
A_\mu^{\mathcal S}
=q\,g_\mu{}^{\bar\alpha}(x,z_p)
\frac{u_{\bar\alpha}}{\bar s}+O(\rho),
\label{eq:em-singular-leading}
\end{equation}
within a convex normal neighborhood of $z_p^\mu$.
Barred indices refer to the reference point $z_p^\mu$,
while unbarred indices refer to the field point $x^\mu$.
Here $u_{\bar\alpha}$ is the covariant four-velocity at $z_p^\mu$,
and $g_\mu{}^{\bar\alpha}(x,z_p)$ is the parallel propagator
from $z_p^\mu$ to $x^\mu$.
The spatial separation in the local rest space at $z_p^\mu$
is defined by
\begin{equation}
\bar s^2=P_{\bar\alpha\bar\beta}
\sigma^{\bar\alpha}\sigma^{\bar\beta},\quad
P_{\bar\alpha\bar\beta}
\equiv g_{\bar\alpha\bar\beta}
+u_{\bar\alpha}u_{\bar\beta}.
\label{eq:sbar-definition}
\end{equation}
Here $\sigma(x,z_p)$ is Synge's world function,
$\sigma^{\bar\alpha}\equiv\nabla^{\bar\alpha}\sigma(x,z_p)$,
and $P_{\bar\alpha\bar\beta}$ projects orthogonally to
$u^{\bar\alpha}$.
Thus, $\bar s=O(\rho)$.

In Schwarzschild coordinates, the world-function gradient expands as
\begin{equation}
\sigma^{\bar\alpha}
=-\Delta x^\alpha
-\frac{1}{2}\left.\Gamma^\alpha{}_{\beta\gamma}\right|_{z_p}
\Delta x^\beta\Delta x^\gamma
+O(\rho^3).
\label{eq:world-function-gradient-expansion}
\end{equation}
Its leading contribution to Eq. \eqref{eq:sbar-definition}
on the equal-time hypersurface is
\begin{equation}
\bar s^2
={}f_p^{-1}(\Delta r)^2
+r_p^2(\Delta\theta)^2
+r_p^2(\Delta\phi)^2
+\left(u_r\Delta r+u_\phi\Delta\phi\right)^2
+O(\rho^3),
\label{eq:sbar-leading-coordinate}
\end{equation}
where $f_p=f(r_p)$ and $u_\theta=0$ for equatorial motion.
All velocity components are evaluated at $z_p^\mu$,
and bars are omitted on explicit coordinate indices.

Equation \eqref{eq:sbar-leading-coordinate} displays only the leading local behavior of $\bar s$ near the particle.
We do not promote this truncated coordinate expansion to a global denominator on the integration sphere.
Instead, the spherical-harmonic construction below uses the analytic scalar puncture modes and their derivatives developed in Refs. \cite{Heffernan:2012su,Heffernan:2017cad,Leather:2023dzj}.

We retain the parallel-transported velocity through first order
in the coordinate separation and define
\begin{equation}
A_\mu^{\mathcal P}
=q\frac{\zeta_\mu}{\bar s},
\label{eq:em-puncture-definition}
\end{equation}
where
\begin{equation}
\zeta_\mu\equiv
\left[g_\mu{}^{\bar\alpha}(x,z_p)
u_{\bar\alpha}\right]_{O(\rho^0)+O(\rho^1)}.
\end{equation}
The required expansion of the parallel propagator is
\begin{equation}
g_\mu{}^{\bar\alpha}(x,z_p)
=\delta_\mu{}^\alpha
+\left.\Gamma^\alpha{}_{\mu\beta}\right|_{z_p}
\Delta x^\beta
+O(\rho^2).
\label{eq:parallel-propagator-expansion}
\end{equation}
Using the Schwarzschild connection coefficients gives
\begin{equation}
\begin{split}
\zeta_t
&=u_t+
\frac{u_t}{r_p(r_p-2)}\Delta r,
\\
\zeta_r
&=u_r
-\frac{u_r}{r_p(r_p-2)}\Delta r
+\frac{u_\phi}{r_p}\Delta\phi,
\\
\zeta_\theta
&=-(r_p-2)u_r\Delta\theta,
\\
\zeta_\phi
&=u_\phi
+\frac{u_\phi}{r_p}\Delta r
-(r_p-2)u_r\Delta\phi.
\end{split}
\label{eq:zeta-components}
\end{equation}
The omitted numerator terms are $O(\rho^2)$.
These coefficients also determine the gravitational puncture numerator $\chi_{\mu\nu}$ used in Ref. \cite{Zhang:2025eqz},
\begin{equation}
\chi_{\mu\nu}
=\left[\zeta_\mu\zeta_\nu\right]_{O(\rho^0)+O(\rho^1)}.
\label{eq:em-gravitational-puncture-relation}
\end{equation}
This relation reflects the single parallel-transported velocity in the electromagnetic singular field and its tensor product in the trace-reversed gravitational singular field.

Together with the scalar puncture construction specified below,
the numerator expansion reproduces the retained singular-field terms, yielding
\begin{equation}
A_\mu^{\mathcal P}-A_\mu^{\mathcal S}=O(\rho).
\label{eq:puncture-local-accuracy}
\end{equation}
The puncture therefore reproduces the $O(\rho^{-1})$
and $O(\rho^0)$ terms of the singular field. Under the convention that
an order-$n$ puncture has error $O(\rho^{n-1})$, this is a second-order
puncture. Its potential error vanishes at the particle, while its
first-derivative error is generally $O(1)$.
The regularity of the resulting effective-source modes
is discussed below.

Before spherical-harmonic projection, the local puncture must
be extended over the integration sphere.
To remove the discontinuity of the azimuthal difference
at $\Delta\phi=\pm\pi$, we use the periodic replacement
\cite{Heffernan:2017cad,Wardell:2015ada,Dolan:2012jg,Zhang:2025eqz}
\begin{equation}
\Delta\phi\longrightarrow\sin\Delta\phi
=\Delta\phi+O(\Delta\phi^3).
\label{eq:azimuthal-completion}
\end{equation}
Similarly, the polar separation can be replaced by
\begin{equation}
\Delta\theta\longrightarrow\sin\Delta\theta
=-\cos\theta
=\Delta\theta+O(\Delta\theta^3),
\label{eq:polar-completion}
\end{equation}
where $\Delta\theta=\theta-\pi/2$.
These replacements preserve the linear angular terms
in the puncture numerator.
The full angular completion introduced below includes
additional angular factors and preserves the puncture
through $O(\rho^0)$.

\section{Analytic construction of the puncture modes}

We now express the vector-harmonic puncture modes as finite linear combinations of the scalar puncture modes $\psi^{\ell m}$.
We first decompose the puncture in a tetrad basis, reduce each tetrad component to scalar spherical harmonics, and then reconstruct the vector-harmonic amplitudes.

\subsection{Tetrad decomposition}

Direct projection of $A_\mu^{\mathcal P}$ onto vector spherical harmonics leads to cumbersome angular integrals.
We instead project the covector field onto a tetrad adapted to the spherical symmetry of the Schwarzschild background.
Now,
we choose the complex tetrad
\begin{equation}
\label{eq:contravariant-tetrad}
\begin{split}
e_{(0)}^\alpha&=\left[\frac{1}{\sqrt{f}},0,0,0\right],
\\
e_{(1)}^\alpha&=e^{i\phi}\left[0,\sqrt{f}\sin\theta,\frac{\cos\theta}{r},\frac{i}{r\sin\theta}\right],
\\
e_{(2)}^\alpha&=e^{-i\phi}\left[0,\sqrt{f}\sin\theta,\frac{\cos\theta}{r},-\frac{i}{r\sin\theta}\right],
\\
e_{(3)}^\alpha&=\left[0,\sqrt{f}\cos\theta,-\frac{\sin\theta}{r},0\right].
\end{split}
\end{equation}
The corresponding dual tetrad is
\begin{equation}
\begin{split}
e_{\alpha}^{(0)}&=\left[\sqrt{f},0,0,0\right],
\\
e_{\alpha}^{(1)}&=\frac{1}{2}e^{-i\phi}\left[0,\frac{\sin\theta}{\sqrt{f}},r\cos\theta,-ir\sin\theta\right],
\\
e_{\alpha}^{(2)}&=\frac{1}{2}e^{i\phi}\left[0,\frac{\sin\theta}{\sqrt{f}},r\cos\theta,ir\sin\theta\right],
\\
e_{\alpha}^{(3)}&=\left[0,\frac{\cos\theta}{\sqrt{f}},-r\sin\theta,0\right].
\end{split}
\end{equation}
Thus,
the tetrad components of the puncture are
\begin{equation}
\label{eq:tetrad-puncture-definition}
A_{(\mu)}^{\mathcal P}
\equiv e_{(\mu)}^\alpha A_\alpha^{\mathcal P}
=\frac{q}{\bar s}e_{(\mu)}^\alpha\zeta_\alpha.
\end{equation}
Equation \eqref{eq:contravariant-tetrad} gives
\begin{equation}
\label{eq:tetrad-puncture-compact}
\begin{split}
A_{(0)}^{\mathcal P}
&=\frac{q}{\bar s}\frac{\zeta_t}{\sqrt{f}},
\\
A_{(1)}^{\mathcal P}
&=\frac{q}{\bar s}e^{i\phi}
\left(
\sqrt{f}\sin\theta\,\zeta_r
+\frac{\cos\theta}{r}\zeta_\theta
+\frac{i}{r\sin\theta}\zeta_\phi
\right),
\\
A_{(2)}^{\mathcal P}
&=\frac{q}{\bar s}e^{-i\phi}
\left(
\sqrt{f}\sin\theta\,\zeta_r
+\frac{\cos\theta}{r}\zeta_\theta
-\frac{i}{r\sin\theta}\zeta_\phi
\right),
\\
A_{(3)}^{\mathcal P}
&=\frac{q}{\bar s}
\left(
\sqrt{f}\cos\theta\,\zeta_r
-\frac{\sin\theta}{r}\zeta_\theta
\right).
\end{split}
\end{equation}
Substituting Eq. \eqref{eq:zeta-components} yields
\begin{equation}
A_{(0)}^{\mathcal P}=\frac{q u_t}{\bar s\sqrt{f}}\left(1+\frac{\Delta r}{r_p(r_p-2)}\right),
\end{equation}

\begin{equation}
\begin{split}
A_{(1)}^{\mathcal P}=\frac{q e^{i\phi}}{\bar s}\Bigg\{&\sqrt{f}\sin\theta\left[u_r-\frac{u_r}{r_p(r_p-2)}\Delta r+\frac{u_\phi}{r_p}\Delta\phi\right]
\\
&
-\frac{(r_p-2)u_r}{r}\cos\theta\,\Delta\theta
\\
&
+\frac{i}{r\sin\theta}\left[u_\phi+\frac{u_\phi}{r_p}\Delta r-(r_p-2)u_r\Delta\phi\right]
\Bigg\},
\end{split}
\end{equation}
\begin{equation}
\begin{split}
A_{(2)}^{\mathcal P}=\frac{q e^{-i\phi}}{\bar s}\Bigg\{&\sqrt{f}\sin\theta\left[u_r-\frac{u_r}{r_p(r_p-2)}\Delta r+\frac{u_\phi}{r_p}\Delta\phi\right]
\\
&
-\frac{(r_p-2)u_r}{r}\cos\theta\,\Delta\theta
\\
&
-\frac{i}{r\sin\theta}\left[u_\phi+\frac{u_\phi}{r_p}\Delta r-(r_p-2)u_r\Delta\phi\right]
\Bigg\}, 
\end{split}
\end{equation}
\begin{equation}
\begin{split}
A_{(3)}^{\mathcal P}=\frac{q}{\bar s}
\Bigg\{&\sqrt{f}\cos\theta\left[u_r-\frac{u_r}{r_p(r_p-2)}\Delta r+\frac{u_\phi}{r_p}\Delta\phi\right]
\\
&
+\frac{(r_p-2)u_r}{r}\sin\theta\,\Delta\theta
\Bigg\}. 
\end{split}
\end{equation}

For fixed $(\mu)$, the tetrad component $A_{(\mu)}^{\mathcal P}$ is a coordinate scalar and may be expanded in ordinary scalar spherical harmonics,
\begin{equation}
A_{(\mu)}^{\mathcal P}(t,r,\theta,\phi)
=\sum_{\ell m}
a_{(\mu)}^{\mathcal P,\ell m}(t,r)
Y_{\ell m}(\theta,\phi),
\label{eq:tetrad-scalar-harmonic-expansion}
\end{equation}
with
\begin{equation}
\label{eq:tetrad-scalar-mode-definition}
a_{(\mu)}^{\mathcal P,\ell m}(t,r)
=\int A_{(\mu)}^{\mathcal P}
Y_{\ell m}^*\,d\Omega.
\end{equation}
We define the scalar puncture modes by
\begin{equation}
\psi^{\ell m}(t,r)
\equiv
r\int\frac{1}{\bar s}Y_{\ell m}^*\,d\Omega.
\label{eq:scalar-puncture-modes}
\end{equation}
The temporal tetrad mode is then
\begin{equation}
a_{(0)}^{\mathcal P,\ell m}
=\frac{q u_t}{r\sqrt{f}}
\left(
1+\frac{\Delta r}{r_p(r_p-2)}
\right)
\psi^{\ell m}.
\label{eq:temporal-tetrad-mode}
\end{equation}
The spatial components contain additional angular factors and require an angular extension of the local puncture.

\subsection{Angular completion and scalar-mode reduction}

We extend the numerator in Eq. \eqref{eq:zeta-components}
over the integration sphere by choosing
\begin{equation}
\label{eq:completed-zeta}
\begin{split}
\hat\zeta_t
&=u_t\left(1+\frac{\Delta r}{r_p(r_p-2)}\right),
\\
\hat\zeta_r
&=u_r\left(1-\frac{\Delta r}{r_p(r_p-2)}\right)
+\frac{u_\phi}{r_p}\sin\theta\sin(\phi-\phi_p),
\\
\hat\zeta_\theta
&=(r_p-2)u_r\sin\theta\cos\theta,
\\
\hat\zeta_\phi
&=u_\phi\left(1+\frac{\Delta r}{r_p}\right)\sin^2\theta
-(r_p-2)u_r\sin^3\theta\sin(\phi-\phi_p).
\end{split}
\end{equation}
A hat denotes the angularly completed numerator.
Since
\begin{equation}
\begin{split}
\sin\theta
&=1+O(\Delta\theta^2),
\\
-\cos\theta
&=\Delta\theta+O(\Delta\theta^3),
\\
\sin(\phi-\phi_p)
&=\Delta\phi+O(\Delta\phi^3),
\end{split}
\end{equation}
we have
$\hat\zeta_\mu-\zeta_\mu=O(\rho^2)$.
Consequently,
\begin{equation}
\hat A_\mu^{\mathcal P}-A_\mu^{\mathcal P}=O(\rho),
\qquad
\hat A_\mu^{\mathcal P}
=q\frac{\hat\zeta_\mu}{\bar s},
\end{equation}
and the retained local singular structure is unchanged.
The angular factors in Eq. \eqref{eq:completed-zeta} also remove the
coordinate singularities arising from the spatial tetrad projections.

For the first spatial tetrad component,
Eq. \eqref{eq:completed-zeta} gives
\begin{equation}
\label{eq:direct-integral-1}
\begin{split}
\hat a_{(1)}^{\mathcal P,\ell m}
&=\int\hat A_{(1)}^{\mathcal P}Y_{\ell m}^*\,d\Omega\\
&=q\int\frac{Y_{\ell m}^*}{\bar s}
\Bigg\{
\left[
\sqrt f\,u_r\left(1-\frac{\Delta r}{r_p(r_p-2)}\right)
+\frac{i u_\phi}{r}\left(1+\frac{\Delta r}{r_p}\right)
\right]\sin\theta e^{i\phi}\\
&\quad
+\frac{(r_p-2)u_r}{r}\sin\theta\cos^2\theta e^{i\phi}\\
&\quad
+\frac{1}{2i}
\left[\frac{\sqrt f\,u_\phi}{r_p}
-\frac{i(r_p-2)u_r}{r}\right]
\left[e^{-i\phi_p}\sin^2\theta e^{2i\phi}
-e^{i\phi_p}\sin^2\theta\right]
\Bigg\}\,d\Omega.
\end{split}
\end{equation}
The angular factors are reduced using the recurrence relations
\begin{equation}
\label{eq:direct-harmonic-recurrences}
\begin{split}
\cos\theta\,Y_{\ell m}^*
&=\sqrt{\frac{(\ell-m+1)(\ell+m+1)}
{(2\ell+1)(2\ell+3)}}Y_{\ell+1,m}^*
+\sqrt{\frac{(\ell-m)(\ell+m)}
{(2\ell-1)(2\ell+1)}}Y_{\ell-1,m}^*,\\
\sin\theta e^{i\phi}Y_{\ell m}^*
&=\sqrt{\frac{(\ell-m+1)(\ell-m+2)}
{(2\ell+1)(2\ell+3)}}Y_{\ell+1,m-1}^*
-\sqrt{\frac{(\ell+m)(\ell+m-1)}
{(2\ell-1)(2\ell+1)}}Y_{\ell-1,m-1}^*,\\
\sin\theta e^{-i\phi}Y_{\ell m}^*
&=-\sqrt{\frac{(\ell+m+1)(\ell+m+2)}
{(2\ell+1)(2\ell+3)}}Y_{\ell+1,m+1}^*
+\sqrt{\frac{(\ell-m)(\ell-m-1)}
{(2\ell-1)(2\ell+1)}}Y_{\ell-1,m+1}^*.
\end{split}
\end{equation}
For example,
\begin{equation}
\label{eq:direct-first-integral}
\begin{split}
\int\frac{\sin\theta e^{i\phi}}{\bar s}
Y_{\ell m}^*\,d\Omega
&=
\sqrt{\frac{(\ell-m+1)(\ell-m+2)}
{(2\ell+1)(2\ell+3)}}
\int\frac{Y_{\ell+1,m-1}^*}{\bar s}\,d\Omega\\
&\quad-
\sqrt{\frac{(\ell+m)(\ell+m-1)}
{(2\ell-1)(2\ell+1)}}
\int\frac{Y_{\ell-1,m-1}^*}{\bar s}\,d\Omega\\
&=\frac{1}{r}
\left[
\sqrt{\frac{(\ell-m+1)(\ell-m+2)}
{(2\ell+1)(2\ell+3)}}\psi^{\ell+1,m-1}
\right.\\
&\quad\left.-
\sqrt{\frac{(\ell+m)(\ell+m-1)}
{(2\ell-1)(2\ell+1)}}\psi^{\ell-1,m-1}
\right].
\end{split}
\end{equation}
Applying the recurrence twice yields
\begin{equation}
\label{eq:direct-second-integral}
\begin{split}
\int\frac{\sin^2\theta e^{2i\phi}}{\bar s}Y_{\ell m}^*\,d\Omega
=\frac{1}{r}\Bigg[
&\frac{\sqrt{(\ell-m+1)(\ell-m+2)(\ell-m+3)(\ell-m+4)}}
{(2\ell+3)\sqrt{(2\ell+1)(2\ell+5)}}\psi^{\ell+2,m-2}\\
&-\frac{2\sqrt{(\ell+m)(\ell+m-1)(\ell-m+1)(\ell-m+2)}}
{(2\ell-1)(2\ell+3)}\psi^{\ell,m-2}\\
&+\frac{\sqrt{(\ell+m)(\ell+m-1)(\ell+m-2)(\ell+m-3)}}
{(2\ell-1)\sqrt{(2\ell-3)(2\ell+1)}}\psi^{\ell-2,m-2}
\Bigg].
\end{split}
\end{equation}
Using $\sin^2\theta=1-\cos^2\theta$ to reduce the remaining
terms, we obtain
\begin{equation}
\label{eq:direct-mode-1}
\begin{split}
\hat a_{(1)}^{\mathcal P,\ell m}=\frac{q}{r}\Bigg[
&\sum_{i=-3,-1,1,3}\kappa^{(1)}_{i,-1}\psi^{\ell+i,m-1}\\
&+\sum_{i=-2,0,2}
\left(\kappa^{(1)}_{i,-2}\psi^{\ell+i,m-2}
+\kappa^{(1)}_{i,0}\psi^{\ell+i,m}\right)
\Bigg].
\end{split}
\end{equation}

The second component gives
\begin{equation}
\label{eq:direct-integral-2}
\begin{split}
\hat a_{(2)}^{\mathcal P,\ell m}
&=\int\hat A_{(2)}^{\mathcal P}Y_{\ell m}^*\,d\Omega\\
&=q\int\frac{Y_{\ell m}^*}{\bar s}
\Bigg\{
\left[
\sqrt f\,u_r\left(1-\frac{\Delta r}{r_p(r_p-2)}\right)
-\frac{i u_\phi}{r}\left(1+\frac{\Delta r}{r_p}\right)
\right]\sin\theta e^{-i\phi}\\
&\quad
+\frac{(r_p-2)u_r}{r}\sin\theta\cos^2\theta e^{-i\phi}\\
&\quad
+\frac{1}{2i}
\left[\frac{\sqrt f\,u_\phi}{r_p}
+\frac{i(r_p-2)u_r}{r}\right]
\left[e^{-i\phi_p}\sin^2\theta
-e^{i\phi_p}\sin^2\theta e^{-2i\phi}\right]
\Bigg\}\,d\Omega.
\end{split}
\end{equation}
The same procedure yields
\begin{equation}
\label{eq:direct-mode-2}
\begin{split}
\hat a_{(2)}^{\mathcal P,\ell m}=\frac{q}{r}\Bigg[
&\sum_{i=-3,-1,1,3}\kappa^{(2)}_{i,1}\psi^{\ell+i,m+1}\\
&+\sum_{i=-2,0,2}
\left(\kappa^{(2)}_{i,2}\psi^{\ell+i,m+2}
+\kappa^{(2)}_{i,0}\psi^{\ell+i,m}\right)
\Bigg].
\end{split}
\end{equation}
For real particle data,
$\hat A_{(2)}^{\mathcal P}=(\hat A_{(1)}^{\mathcal P})^*$
implies the mode relation
\begin{equation}
\label{eq:direct-mode-conjugacy}
\hat a_{(2)}^{\mathcal P,\ell m}
=(-1)^m\left[\hat a_{(1)}^{\mathcal P,\ell,-m}\right]^*.
\end{equation}

For the third component,
\begin{equation}
\label{eq:direct-integral-3}
\begin{split}
\hat a_{(3)}^{\mathcal P,\ell m}
&=\int\hat A_{(3)}^{\mathcal P}Y_{\ell m}^*\,d\Omega\\
&=q\int\frac{Y_{\ell m}^*}{\bar s}
\Bigg\{
\left[
\sqrt f\,u_r\left(1-\frac{\Delta r}{r_p(r_p-2)}\right)
-\frac{(r_p-2)u_r}{r}
\right]\cos\theta\\
&\quad+\frac{(r_p-2)u_r}{r}\cos^3\theta\\
&\quad+\frac{\sqrt f\,u_\phi}{2i r_p}
\left[e^{-i\phi_p}\sin\theta\cos\theta e^{i\phi}
-e^{i\phi_p}\sin\theta\cos\theta e^{-i\phi}\right]
\Bigg\}\,d\Omega.
\end{split}
\end{equation}
Repeated application of Eq. \eqref{eq:direct-harmonic-recurrences}
gives
\begin{equation}
\label{eq:direct-mode-3}
\begin{split}
\hat a_{(3)}^{\mathcal P,\ell m}=\frac{q}{r}\Bigg[
&\sum_{i=-3,-1,1,3}\kappa^{(3)}_{i,0}\psi^{\ell+i,m}\\
&+\sum_{i=-2,0,2}
\left(\kappa^{(3)}_{i,-1}\psi^{\ell+i,m-1}
+\kappa^{(3)}_{i,1}\psi^{\ell+i,m+1}\right)
\Bigg].
\end{split}
\end{equation}

All three spatial components therefore take the finite-sum form
\begin{equation}
\label{eq:electromagnetic-finite-mode-sum}
\begin{split}
\hat a_{(\mu)}^{\mathcal P,\ell m}
&=\int\hat A_{(\mu)}^{\mathcal P}Y_{\ell m}^*\,d\Omega\\
&=q\int\frac{1}{\bar s}
\sum_{i,j}\kappa^{(\mu)}_{ij}
Y_{\ell+i,m+j}^*\,d\Omega\\
&=\frac{q}{r}\sum_{i,j}\kappa^{(\mu)}_{ij}
\psi^{\ell+i,m+j}.
\end{split}
\end{equation}
The coefficients $\kappa^{(\mu)}_{ij}(t,r;\ell,m)$
are listed in Appendix \ref{app:direct-mode-coefficients}.
Only the shifts displayed above contribute, and terms with
$\ell+i<0$ or $|m+j|>\ell+i$ are omitted.
For the scalar modes $\psi^{\ell m}$, we use the analytic
eccentric-orbit puncture modes developed in
Refs. \cite{Heffernan:2012su,Heffernan:2017cad,Leather:2023dzj}
and employed in Ref. \cite{Zhang:2025eqz}.
Explicit expressions for $\psi^{\ell m}$ and its first and second
radial and temporal derivatives are provided in the supplemental
Mathematica notebook accompanying Ref. \cite{Leather:2023dzj}.

\begin{table*}[!tp]
\caption{
Nonvanishing tetrad-to-vector coupling coefficients
$C^{(i)(\mu)}_{\ell m\,|\,\ell'm'}$, with
$\lambda=\ell(\ell+1)$.
The indices $i=1,2,3,4$ denote the temporal, radial, even-parity angular, and odd-parity angular components, respectively.
All unlisted coefficients vanish.
}
\label{tab:em-C-coefficients}
\begingroup
\small
\renewcommand{\arraystretch}{0.9}
\setlength{\tabcolsep}{8pt}
\setlength{\jot}{1pt}
\begin{ruledtabular}
\begin{tabular}{c@{\hspace{0.7cm}}l}
$(i,\mu)$
&
\multicolumn{1}{c}{Nonvanishing coefficients}
\\
\hline
$(1,0)$
&
$\displaystyle
C^{(1)(0)}_{\ell m\,|\,\ell m}
=\sqrt f
$
\\
\hline
$(2,1)$
&
$\begin{aligned}
C^{(2)(1)}_{\ell m\,|\,\ell+1,m+1}
&=-\frac{\sqrt f}{2}
\sqrt{\frac{(\ell+m+1)(\ell+m+2)}
{(2\ell+1)(2\ell+3)}}
\\
C^{(2)(1)}_{\ell m\,|\,\ell-1,m+1}
&=\frac{\sqrt f}{2}
\sqrt{\frac{(\ell-m)(\ell-m-1)}
{(2\ell-1)(2\ell+1)}}
\end{aligned}$
\\
$(2,2)$
&
$\begin{aligned}
C^{(2)(2)}_{\ell m\,|\,\ell+1,m-1}
&=\frac{\sqrt f}{2}
\sqrt{\frac{(\ell-m+1)(\ell-m+2)}
{(2\ell+1)(2\ell+3)}}
\\
C^{(2)(2)}_{\ell m\,|\,\ell-1,m-1}
&=-\frac{\sqrt f}{2}
\sqrt{\frac{(\ell+m)(\ell+m-1)}
{(2\ell-1)(2\ell+1)}}
\end{aligned}$
\\
$(2,3)$
&
$\begin{aligned}
C^{(2)(3)}_{\ell m\,|\,\ell+1,m}
&=\sqrt f
\sqrt{\frac{(\ell-m+1)(\ell+m+1)}
{(2\ell+1)(2\ell+3)}}
\\
C^{(2)(3)}_{\ell m\,|\,\ell-1,m}
&=\sqrt f
\sqrt{\frac{(\ell-m)(\ell+m)}
{(2\ell-1)(2\ell+1)}}
\end{aligned}$
\\
\hline
$(3,1)$
&
$\begin{aligned}
C^{(3)(1)}_{\ell m\,|\,\ell+1,m+1}
&=\frac{\ell}{2\sqrt\lambda}
\sqrt{\frac{(\ell+m+1)(\ell+m+2)}
{(2\ell+1)(2\ell+3)}}
\\
C^{(3)(1)}_{\ell m\,|\,\ell-1,m+1}
&=\frac{\ell+1}{2\sqrt\lambda}
\sqrt{\frac{(\ell-m)(\ell-m-1)}
{(2\ell-1)(2\ell+1)}}
\end{aligned}$
\\

$(3,2)$
&
$\begin{aligned}
C^{(3)(2)}_{\ell m\,|\,\ell+1,m-1}
&=-\frac{\ell}{2\sqrt\lambda}
\sqrt{\frac{(\ell-m+1)(\ell-m+2)}
{(2\ell+1)(2\ell+3)}}
\\
C^{(3)(2)}_{\ell m\,|\,\ell-1,m-1}
&=-\frac{\ell+1}{2\sqrt\lambda}
\sqrt{\frac{(\ell+m)(\ell+m-1)}
{(2\ell-1)(2\ell+1)}}
\end{aligned}$
\\
$(3,3)$
&
$\begin{aligned}
C^{(3)(3)}_{\ell m\,|\,\ell+1,m}
&=-\frac{\ell}{\sqrt\lambda}
\sqrt{\frac{(\ell-m+1)(\ell+m+1)}
{(2\ell+1)(2\ell+3)}}
\\
C^{(3)(3)}_{\ell m\,|\,\ell-1,m}
&=\frac{\ell+1}{\sqrt\lambda}
\sqrt{\frac{(\ell-m)(\ell+m)}
{(2\ell-1)(2\ell+1)}}
\end{aligned}$
\\
\hline
$(4,1)$
&
$
C^{(4)(1)}_{\ell m\,|\,\ell,m+1}
=\frac{i}{2\sqrt\lambda}
\sqrt{(\ell-m)(\ell+m+1)}
$
\\
$(4,2)$
&
$
C^{(4)(2)}_{\ell m\,|\,\ell,m-1}
=\frac{i}{2\sqrt\lambda}
\sqrt{(\ell+m)(\ell-m+1)}
$
\\
$(4,3)$
&
$
C^{(4)(3)}_{\ell m\,|\,\ell m}
=\frac{im}{\sqrt\lambda}
$
\end{tabular}
\end{ruledtabular}
\endgroup
\end{table*}

\subsection{Reconstruction of vector-harmonic puncture modes}
\label{sec:tetrad-vector-transformation}

Combining the dual-tetrad reconstruction with the vector-harmonic projection gives
\begin{equation}
\label{eq:em-C-transformation}
\mathcal A_{\ell m}^{(i),\mathcal P}
=\frac{r}{c_i}
\sum_{\mu=0}^{3}\sum_{\ell'm'}
C^{(i)(\mu)}_{\ell m\,|\,\ell'm'}
\hat a_{(\mu)}^{\mathcal P,\ell'm'},
\end{equation}
where
\begin{equation}
\label{eq:em-C-definition}
C^{(i)(\mu)}_{\ell m\,|\,\ell'm'}
\equiv
\int
\bigl(Z_\alpha^{(i)\ell m}\bigr)^*
\eta^{\alpha\beta}e_\beta^{(\mu)}
Y_{\ell'm'}\,d\Omega.
\end{equation}
The recurrence relations in Eq. \eqref{eq:direct-harmonic-recurrences} and spherical-harmonic orthogonality yield the coefficients listed in Table \ref{tab:em-C-coefficients}.
For $\ell=0$, only the temporal and radial harmonics are present,
and the coefficients with $i=3,4$ apply for $\ell\geq1$.

Substituting Eq. \eqref{eq:electromagnetic-finite-mode-sum}
into Eq. \eqref{eq:em-C-transformation} gives
\begin{equation}
\label{eq:em-vector-puncture-scalar-composition}
\mathcal A_{\ell m}^{(i),\mathcal P}
=\frac{q}{c_i}
\sum_{\mu=0}^{3}\sum_{\ell'm'}\sum_{p,s}
C^{(i)(\mu)}_{\ell m\,|\,\ell'm'}
\kappa^{(\mu)}_{ps}(t,r;\ell',m')
\psi^{\ell'+p,m'+s}.
\end{equation}
Here $p$ and $s$ denote harmonic-index shifts.
The temporal component is included by defining
\begin{equation}
\kappa^{(0)}_{00}
=\frac{u_t}{\sqrt f}
\left(1+\frac{\Delta r}{r_p(r_p-2)}\right),
\end{equation}
with all other $\kappa^{(0)}_{ps}$ vanishing.
Since both $C$ and $\kappa$ have finite support,
Eq. \eqref{eq:em-vector-puncture-scalar-composition} expresses every vector-harmonic puncture mode as a finite sum of scalar modes, ready for application of the Maxwell mode operators.

\section{Electromagnetic effective-source modes}
\label{subsec:effective-source-formulation}

The vector-harmonic puncture modes are given by Eq. \eqref{eq:em-vector-puncture-scalar-composition}.
We substitute them into the Maxwell mode equations to obtain the effective sources.
The residual modes obey the same differential operators as the retarded modes, with source terms
\begin{equation}
\begin{split}
S_{\ell m}^{(1),\mathrm{eff}}
={}&-4\pi f\mathcal J_{\ell m}^{(1)}
-\mathcal D_\ell\mathcal A_{\ell m}^{(1),\mathcal P}
-\frac{2}{r^2}\left(
\partial_t\mathcal A_{\ell m}^{(2),\mathcal P}
-\partial_{r_*}\mathcal A_{\ell m}^{(1),\mathcal P}
\right),
\\
S_{\ell m}^{(2),\mathrm{eff}}
={}&-4\pi f\mathcal J_{\ell m}^{(2)}
-\mathcal D_\ell\mathcal A_{\ell m}^{(2),\mathcal P}
\\
&-\frac{2}{r^2}\left[
\partial_t\mathcal A_{\ell m}^{(1),\mathcal P}
-\partial_{r_*}\mathcal A_{\ell m}^{(2),\mathcal P}
-f^2\left(
\mathcal A_{\ell m}^{(2),\mathcal P}
-\mathcal A_{\ell m}^{(3),\mathcal P}
\right)
\right],
\\
S_{\ell m}^{(3),\mathrm{eff}}
={}&-4\pi f\mathcal J_{\ell m}^{(3)}
-\mathcal D_\ell\mathcal A_{\ell m}^{(3),\mathcal P}
-\frac{2f\ell(\ell+1)}{r^2}
\mathcal A_{\ell m}^{(2),\mathcal P},
\\
S_{\ell m}^{(4),\mathrm{eff}}
={}&-4\pi f\mathcal J_{\ell m}^{(4)}
-\mathcal D_\ell\mathcal A_{\ell m}^{(4),\mathcal P}.
\end{split}
\label{eq:mode-effective-source-components}
\end{equation}
The first three equations describe the coupled even-parity sector.
The fourth describes the decoupled odd-parity sector.
For $\ell=0$, only the first two equations remain.

Each puncture mode is a finite sum of scalar modes.
We therefore evaluate the derivatives in Eq. \eqref{eq:mode-effective-source-components} term by term.
Radial derivatives act on $C$, $\kappa$, and $\psi^{\ell m}$,
with $\partial_{r_*}=f\partial_r$.
At fixed $r$, the coefficients $C$ are time independent.
Time derivatives act on $\kappa$ and $\psi^{\ell m}$ and include their dependence on the particle position and velocity.
We use the Schwarzschild geodesic equations to evaluate the derivatives of these orbital quantities.
The angular term in $\mathcal D_\ell$ uses the vector-harmonic index $\ell$, which may differ from the scalar-mode index in the finite sum.

For numerical evaluation, we use the scalar-mode expressions and their derivatives from the construction adopted above \cite{Zhang:2025eqz}.
The coefficients are differentiated analytically before the orbital parameters are substituted.
The resulting effective sources are then evaluated at fixed orbital phase as functions of $r$.
No additional angular integrations are required.

For $r\ne r_p(t)$, the current modes vanish.
The effective sources then follow from the puncture terms in Eq. \eqref{eq:mode-effective-source-components}.
At $r=r_p(t)$, derivatives of the puncture generate delta-function terms that cancel the point-particle sources.
This cancellation removes the distributional contribution.
The regularity of the remaining mode sources
depends on the puncture order.
Their behavior at the particle is determined by the limits from $r<r_p$ and $r>r_p$.

We illustrate the construction for an eccentric bound orbit.
The particle radius is parametrized by
\begin{equation}
r_p=\frac{p}{1+e\cos\chi_p},
\label{eq:effective-source-orbit-radius}
\end{equation}
where $p$ is the dimensionless semilatus rectum,
$e$ is the eccentricity, and $\chi_p$ is the radial phase.
We choose $p=10$, $e=0.3$, and $\chi_p=\pi/2$.
The particle is then moving outward at $r_p=10$.
We set $q=1$ and choose the azimuthal origin so that $\phi_p=0$ at the evaluation time.

\begin{figure}[htp]
\centering
\includegraphics[width=0.97\columnwidth]{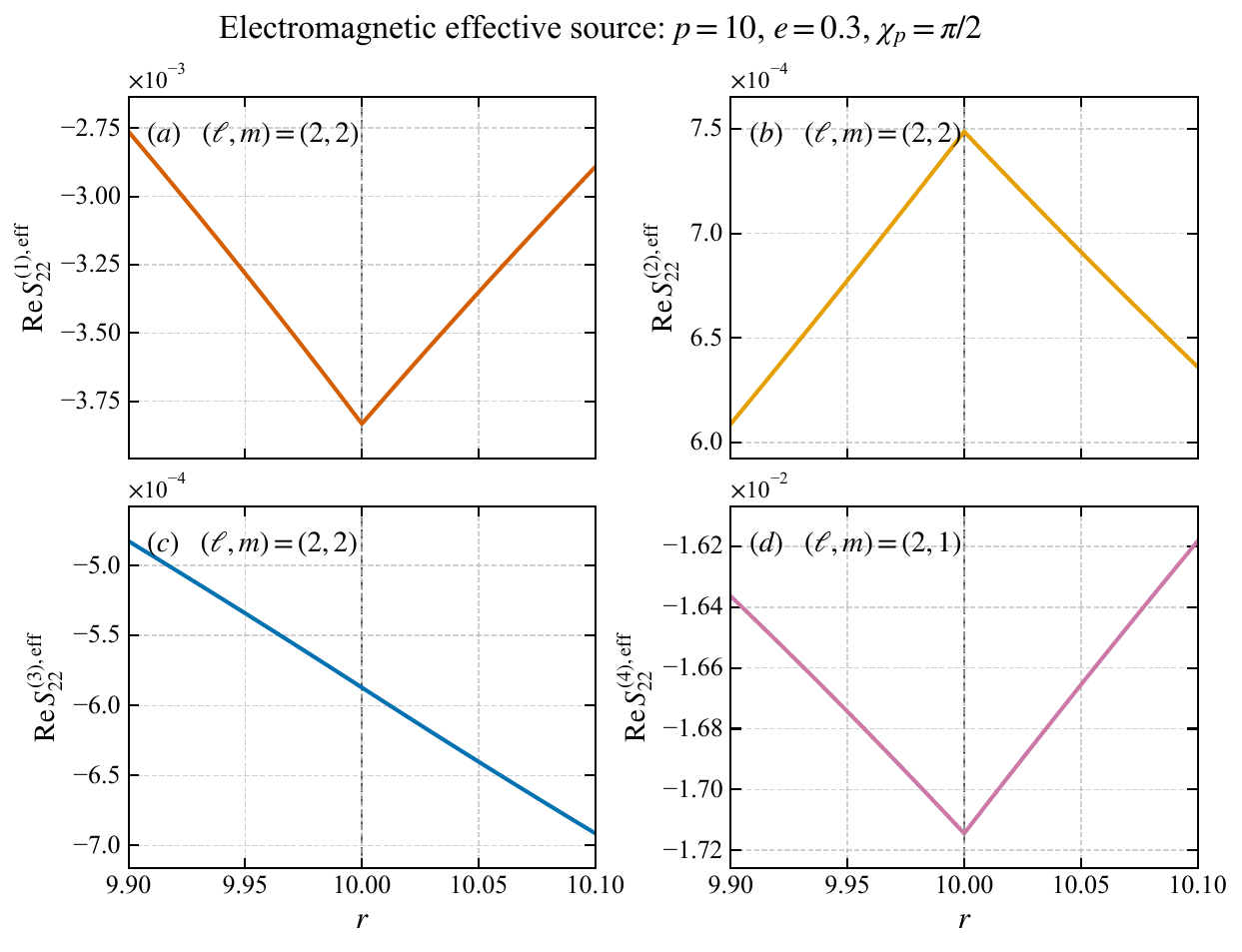}
\caption{
Real parts of the electromagnetic effective-source modes
for an eccentric Schwarzschild orbit with $p=10$, $e=0.3$,
and $\chi_p=\pi/2$.
We set $q=1$ and $\phi_p=0$.
Panels (a)--(c) show the $i=1,2,3$ components
with $(\ell,m)=(2,2)$.
Panel (d) shows the $i=4$ component with $(\ell,m)=(2,1)$.
The vertical dashed line marks the particle position $r_p=10$.
}
\label{fig:effsou}
\end{figure}

Figure \ref{fig:effsou} shows the real parts of representative even- and odd-parity effective-source modes over $9.9\leq r\leq10.1$, with the particle at $r_p=10$.
At this phase of the eccentric orbit, the particle has nonzero radial velocity.
All four profiles remain finite near the particle. 
The temporal, radial, and odd-parity profiles exhibit cusp-like behavior at $r_p$, consistent with the limited smoothness expected from the second-order puncture.
These results illustrate the construction of effective-source modes for both parity sectors under noncircular motion.

\section{Conclusion}

In this paper, we construct analytic electromagnetic effective-source $\ell m$ modes for a point charge following a generic geodesic in Schwarzschild spacetime.
Starting from the local Lorenz-gauge Detweiler--Whiting singular potential, we retain its inverse-distance and constant-order terms in a second-order puncture.
We extend the puncture over the sphere while preserving this local accuracy. 
A complex tetrad then reduces its components to finite sums of previously derived scalar modes. 
Using the angular substitutions introduced by the completion, 
together with spherical-harmonic recurrence and orthogonality relations,
we derive the coupling coefficients that reconstruct the even- and odd-parity vector-harmonic puncture amplitudes.

Applying the Maxwell mode operators to the vector-harmonic puncture amplitudes gives the effective-source modes.
The derivatives act on both the scalar modes and their coefficients, including their dependence on the evolving particle position and velocity.
The distributional terms generated by the puncture cancel the corresponding terms in the point-particle current, and no further angular integrations are required.
Representative eccentric-orbit profiles show that the projected sources remain finite at the particle.
Some exhibit cusp-like radial behavior, consistent with the limited smoothness expected from the second-order puncture used here.
These analytic source terms provide the basis for a subsequent $1+1$-dimensional residual-field calculation.

In the future, we will combine this construction with the extended effective-source method to solve the residual-field equations and compute the electromagnetic self-force on eccentric orbits.
Because the puncture used here is second order, direct force extraction requires a higher-order puncture or an equivalent correction.
We therefore also aim to derive higher-order electromagnetic punctures and their corresponding effective sources.

\begin{acknowledgments}
This research is supported in part by the National Natural Science Foundation of China under Grant Nos.  12535002, 12588101 and 12505076, 
and the Ningbo Science \& Technology Bureau under program grant No. 2024DJRC010009.
\end{acknowledgments}

\appendix

\section{Point-particle source modes}
\label{app:point-particle-source}

We define the angular factors
\begin{equation}
\mathcal{Y}_{\ell m}
\equiv
Y_{\ell m}\left(\frac{\pi}{2},0\right),
\quad
\mathcal{Y}_{\ell m,\theta}
\equiv
\left.
\partial_\theta Y_{\ell m}(\theta,0)
\right|_{\theta=\pi/2}.
\label{eq:angular-harmonic-values}
\end{equation}
Projecting the point-particle current onto the vector-harmonic basis gives
\begin{equation}
\begin{split}
\mathcal{S}_{\ell m}^{(1)}
&=
\frac{4\pi q f_p^2}{r_p}
\mathcal{Y}_{\ell m}^{*}e^{-im\phi_p},
\\
\mathcal{S}_{\ell m}^{(2)}
&=
-\frac{4\pi q f_p}{r_p}
\frac{dr_p}{dt}
\mathcal{Y}_{\ell m}^{*}e^{-im\phi_p},
\\
\mathcal{S}_{\ell m}^{(3)}
&=
4\pi i m q f_p
\frac{d\phi_p}{dt}
\mathcal{Y}_{\ell m}^{*}e^{-im\phi_p},
\\
\mathcal{S}_{\ell m}^{(4)}
&=
4\pi q f_p
\frac{d\phi_p}{dt}
\mathcal{Y}_{\ell m,\theta}^{*}e^{-im\phi_p}.
\end{split}
\label{eq:point-particle-source-modes}
\end{equation}
Here $f_p=f[r_p(t)]$, and all orbital quantities are evaluated on the particle worldline. 

\section{Explicit coefficients for the angular projections}
\label{app:direct-mode-coefficients}

This appendix lists the coefficients appearing in
Eqs. \eqref{eq:direct-mode-1}--\eqref{eq:direct-mode-3}.
All particle quantities are evaluated at the reference time.
Only terms satisfying $\ell+i\geq0$ and $|m+j|\leq\ell+i$ are retained.
The spherical-harmonic phase convention is that of
Eq. \eqref{eq:direct-harmonic-recurrences}.

\subsection{The $A_{(1)}^{\mathcal P}$ component}

The coefficients with $j=-1$ arise from
$\sin\theta e^{i\phi}$ and
$\sin\theta\cos^2\theta e^{i\phi}$:
\begin{equation}
\begin{split}
\kappa^{(1)}_{3,-1}
&=\frac{(r_p-2)u_r}{r}\\
&\quad{}\times \left(\frac{\sqrt{\left(\ell - m + 1\right) \left(\ell - m + 2\right) \left(\ell - m + 3\right) \left(\ell - m + 4\right) \left(\ell + m + 1\right) \left(\ell + m + 2\right)}}{\sqrt{\left(2 \ell + 1\right) \left(2 \ell + 7\right)} \left(2 \ell + 3\right) \left(2 \ell + 5\right)}\right).
\end{split}
\end{equation}
\begin{equation}
\begin{split}
\kappa^{(1)}_{1,-1}
&=\left[\sqrt f\,u_r\left(1-\frac{\Delta r}{r_p(r_p-2)}\right)+\frac{i u_\phi}{r}\left(1+\frac{\Delta r}{r_p}\right)\right]\\
&\quad{}\times \left(\sqrt{\frac{\left(\ell - m + 1\right) \left(\ell - m + 2\right)}{\left(2 \ell + 1\right) \left(2 \ell + 3\right)}}\right)\\
&\quad+\frac{(r_p-2)u_r}{r}\\
&\quad{}\times \left(\frac{\sqrt{\frac{\left(\ell - m + 1\right) \left(\ell - m + 2\right)}{\left(2 \ell + 1\right) \left(2 \ell + 3\right)}} \left(\ell^{2} - 2 \ell m + 3 \ell - 3 m^{2} + m - 1\right)}{\left(2 \ell - 1\right) \left(2 \ell + 5\right)}\right).
\end{split}
\end{equation}
\begin{equation}
\begin{split}
\kappa^{(1)}_{-1,-1}
&=\left[\sqrt f\,u_r\left(1-\frac{\Delta r}{r_p(r_p-2)}\right)+\frac{i u_\phi}{r}\left(1+\frac{\Delta r}{r_p}\right)\right]\\
&\quad{}\times \left(- \sqrt{\frac{\left(\ell + m\right) \left(\ell + m - 1\right)}{\left(2 \ell - 1\right) \left(2 \ell + 1\right)}}\right)\\
&\quad+\frac{(r_p-2)u_r}{r}\\
&\quad{}\times \left(\frac{\sqrt{\frac{\left(\ell + m\right) \left(\ell + m - 1\right)}{\left(2 \ell - 1\right) \left(2 \ell + 1\right)}} \left(- \ell^{2} - 2 \ell m + \ell + 3 m^{2} - 3 m + 3\right)}{\left(2 \ell - 3\right) \left(2 \ell + 3\right)}\right).
\end{split}
\end{equation}
\begin{equation}
\begin{split}
\kappa^{(1)}_{-3,-1}
&=\frac{(r_p-2)u_r}{r}\\
&\quad{}\times \left(- \frac{\sqrt{\left(\ell - m\right) \left(\ell + m\right) \left(\ell - m - 1\right) \left(\ell + m - 3\right) \left(\ell + m - 2\right) \left(\ell + m - 1\right)}}{\sqrt{\left(2 \ell - 5\right) \left(2 \ell + 1\right)} \left(2 \ell - 3\right) \left(2 \ell - 1\right)}\right).
\end{split}
\end{equation}

The coefficients with $j=-2$ arise from
$\sin^2\theta e^{2i\phi}$:
\begin{equation}
\begin{split}
\kappa^{(1)}_{2,-2}
&=\frac{e^{-i\phi_p}}{2i}\left[\frac{\sqrt f\,u_\phi}{r_p}-\frac{i(r_p-2)u_r}{r}\right]\\
&\quad{}\times \left(\frac{\sqrt{\left(\ell - m + 1\right) \left(\ell - m + 2\right) \left(\ell - m + 3\right) \left(\ell - m + 4\right)}}{\sqrt{\left(2 \ell + 1\right) \left(2 \ell + 5\right)} \left(2 \ell + 3\right)}\right).
\end{split}
\end{equation}
\begin{equation}
\begin{split}
\kappa^{(1)}_{0,-2}
&=\frac{e^{-i\phi_p}}{2i}\left[\frac{\sqrt f\,u_\phi}{r_p}-\frac{i(r_p-2)u_r}{r}\right]\\
&\quad{}\times \left(- \frac{2 \sqrt{\left(\ell + m\right) \left(\ell - m + 1\right) \left(\ell - m + 2\right) \left(\ell + m - 1\right)}}{\left(2 \ell - 1\right) \left(2 \ell + 3\right)}\right).
\end{split}
\end{equation}
\begin{equation}
\begin{split}
\kappa^{(1)}_{-2,-2}
&=\frac{e^{-i\phi_p}}{2i}\left[\frac{\sqrt f\,u_\phi}{r_p}-\frac{i(r_p-2)u_r}{r}\right]\\
&\quad{}\times \left(\frac{\sqrt{\left(\ell + m\right) \left(\ell + m - 3\right) \left(\ell + m - 2\right) \left(\ell + m - 1\right)}}{\sqrt{\left(2 \ell - 3\right) \left(2 \ell + 1\right)} \left(2 \ell - 1\right)}\right).
\end{split}
\end{equation}

The coefficients with $j=0$ arise from $\sin^2\theta$:
\begin{equation}
\begin{split}
\kappa^{(1)}_{2,0}
&=-\frac{e^{i\phi_p}}{2i}\left[\frac{\sqrt f\,u_\phi}{r_p}-\frac{i(r_p-2)u_r}{r}\right]\\
&\quad{}\times \left(- \frac{\sqrt{\left[-m^{2}+\left(\ell+1\right)^{2}\right]\left[-m^{2}+\left(\ell+2\right)^{2}\right]}}{\sqrt{\left(2\ell+1\right)\left(2\ell+5\right)}\left(2\ell+3\right)}\right).
\end{split}
\end{equation}
\begin{equation}
\begin{split}
\kappa^{(1)}_{0,0}
&=-\frac{e^{i\phi_p}}{2i}\left[\frac{\sqrt f\,u_\phi}{r_p}-\frac{i(r_p-2)u_r}{r}\right]\\
&\quad{}\times \left(\frac{2\ell\left(\ell+1\right)+2m^{2}-2}{\left(2\ell-1\right)\left(2\ell+3\right)}\right).
\end{split}
\end{equation}
\begin{equation}
\begin{split}
\kappa^{(1)}_{-2,0}
&=-\frac{e^{i\phi_p}}{2i}\left[\frac{\sqrt f\,u_\phi}{r_p}-\frac{i(r_p-2)u_r}{r}\right]\\
&\quad{}\times \left(- \frac{\sqrt{\left(\ell^{2}-m^{2}\right)\left[-m^{2}+\left(\ell-1\right)^{2}\right]}}{\sqrt{\left(2\ell-3\right)\left(2\ell+1\right)}\left(2\ell-1\right)}\right).
\end{split}
\end{equation}

\subsection{The $A_{(2)}^{\mathcal P}$ component}

No independent coefficients are required. Equation \eqref{eq:direct-mode-conjugacy} and
$[\psi^{\ell m}]^*=(-1)^m\psi^{\ell,-m}$ give
\begin{equation}
\begin{split}
\kappa^{(2)}_{i,1}(t,r;\ell,m)
&=-[\kappa^{(1)}_{i,-1}(t,r;\ell,-m)]^*,
\qquad i=-3,-1,1,3,
\\
\kappa^{(2)}_{i,2}(t,r;\ell,m)
&=[\kappa^{(1)}_{i,-2}(t,r;\ell,-m)]^*,
\qquad i=-2,0,2,
\\
\kappa^{(2)}_{i,0}(t,r;\ell,m)
&=[\kappa^{(1)}_{i,0}(t,r;\ell,-m)]^*,
\qquad i=-2,0,2.
\end{split}
\end{equation}
Thus the $A_{(2)}^{\mathcal P}$ coefficients follow by replacing
$m$ with $-m$ and taking the complex conjugate, with an additional minus sign for the $j=1$ coefficients.

\subsection{The $A_{(3)}^{\mathcal P}$ component}

The coefficients with $j=0$ arise from
$\cos\theta$ and $\cos^3\theta$:
\begin{equation}
\begin{split}
\kappa^{(3)}_{3,0}
&=\frac{(r_p-2)u_r}{r}\\
&\quad{}\times \left(\frac{\sqrt{\left[-m^{2}+\left(\ell+1\right)^{2}\right]\left[-m^{2}+\left(\ell+2\right)^{2}\right]\left[-m^{2}+\left(\ell+3\right)^{2}\right]}}{\sqrt{\left(2\ell+1\right)\left(2\ell+7\right)}\left(2\ell+3\right)\left(2\ell+5\right)}\right).
\end{split}
\end{equation}
\begin{equation}
\begin{split}
\kappa^{(3)}_{1,0}
&=\left[\sqrt f\,u_r\left(1-\frac{\Delta r}{r_p(r_p-2)}\right)-\frac{(r_p-2)u_r}{r}\right]\\
&\quad{}\times \left(\sqrt{\frac{-m^{2}+\left(\ell+1\right)^{2}}{\left(2\ell+1\right)\left(2\ell+3\right)}}\right)\\
&\quad+\frac{(r_p-2)u_r}{r}\\
&\quad{}\times \left(\frac{3\sqrt{\frac{-m^{2}+\left(\ell+1\right)^{2}}{\left(2\ell+1\right)\left(2\ell+3\right)}}\left(\ell^{2}+2\ell-m^{2}-1\right)}{\left(2\ell-1\right)\left(2\ell+5\right)}\right).
\end{split}
\end{equation}
\begin{equation}
\begin{split}
\kappa^{(3)}_{-1,0}
&=\left[\sqrt f\,u_r\left(1-\frac{\Delta r}{r_p(r_p-2)}\right)-\frac{(r_p-2)u_r}{r}\right]\\
&\quad{}\times \left(\sqrt{\frac{\ell^{2}-m^{2}}{\left(2\ell-1\right)\left(2\ell+1\right)}}\right)\\
&\quad+\frac{(r_p-2)u_r}{r}\\
&\quad{}\times \left(\frac{3\sqrt{\frac{\ell^{2}-m^{2}}{\left(2\ell-1\right)\left(2\ell+1\right)}}\left(\ell^{2}-m^{2}-2\right)}{\left(2\ell-3\right)\left(2\ell+3\right)}\right).
\end{split}
\end{equation}
\begin{equation}
\begin{split}
\kappa^{(3)}_{-3,0}
&=\frac{(r_p-2)u_r}{r}\\
&\quad{}\times \left(\frac{\sqrt{\left(\ell^{2}-m^{2}\right)\left[-m^{2}+\left(\ell-2\right)^{2}\right]\left[-m^{2}+\left(\ell-1\right)^{2}\right]}}{\sqrt{\left(2\ell-5\right)\left(2\ell+1\right)}\left(2\ell-3\right)\left(2\ell-1\right)}\right).
\end{split}
\end{equation}

The coefficients with $j=-1$ arise from
$\sin\theta\cos\theta e^{i\phi}$:
\begin{equation}
\begin{split}
\kappa^{(3)}_{2,-1}
&=\frac{\sqrt f\,u_\phi e^{-i\phi_p}}{2i r_p}\\
&\quad{}\times \left(\frac{\sqrt{\left(\ell-m+1\right)\left(\ell-m+2\right)\left(\ell-m+3\right)\left(\ell+m+1\right)}}{\sqrt{\left(2\ell+1\right)\left(2\ell+5\right)}\left(2\ell+3\right)}\right).
\end{split}
\end{equation}
\begin{equation}
\begin{split}
\kappa^{(3)}_{0,-1}
&=\frac{\sqrt f\,u_\phi e^{-i\phi_p}}{2i r_p}\\
&\quad{}\times \left(\frac{\sqrt{\left(\ell+m\right)\left(\ell-m+1\right)}\left(1-2m\right)}{\left(2\ell-1\right)\left(2\ell+3\right)}\right).
\end{split}
\end{equation}
\begin{equation}
\begin{split}
\kappa^{(3)}_{-2,-1}
&=\frac{\sqrt f\,u_\phi e^{-i\phi_p}}{2i r_p}\\
&\quad{}\times \left(- \frac{\sqrt{\left(\ell-m\right)\left(\ell+m\right)\left(\ell+m-2\right)\left(\ell+m-1\right)}}{\sqrt{\left(2\ell-3\right)\left(2\ell+1\right)}\left(2\ell-1\right)}\right).
\end{split}
\end{equation}

The coefficients with $j=1$ follow from the reality of
$\hat A_{(3)}^{\mathcal P}$:
\begin{equation}
\kappa^{(3)}_{i,1}(t,r;\ell,m)
=-[\kappa^{(3)}_{i,-1}(t,r;\ell,-m)]^*,
\quad i=-2,0,2.
\end{equation}
This relation replaces $m$ by $-m$ and
$e^{-i\phi_p}$ by $e^{i\phi_p}$ while leaving the factor $1/(2i)$ unchanged.
All coefficients not listed above vanish.

%


\begin{thebibliography}{49}%
\makeatletter
\providecommand \@ifxundefined [1]{%
 \@ifx{#1\undefined}
}%
\providecommand \@ifnum [1]{%
 \ifnum #1\expandafter \@firstoftwo
 \else \expandafter \@secondoftwo
 \fi
}%
\providecommand \@ifx [1]{%
 \ifx #1\expandafter \@firstoftwo
 \else \expandafter \@secondoftwo
 \fi
}%
\providecommand \natexlab [1]{#1}%
\providecommand \enquote  [1]{``#1''}%
\providecommand \bibnamefont  [1]{#1}%
\providecommand \bibfnamefont [1]{#1}%
\providecommand \citenamefont [1]{#1}%
\providecommand \href@noop [0]{\@secondoftwo}%
\providecommand \href [0]{\begingroup \@sanitize@url \@href}%
\providecommand \@href[1]{\@@startlink{#1}\@@href}%
\providecommand \@@href[1]{\endgroup#1\@@endlink}%
\providecommand \@sanitize@url [0]{\catcode `\\12\catcode `\$12\catcode
  `\&12\catcode `\#12\catcode `\^12\catcode `\_12\catcode `\%12\relax}%
\providecommand \@@startlink[1]{}%
\providecommand \@@endlink[0]{}%
\providecommand \url  [0]{\begingroup\@sanitize@url \@url }%
\providecommand \@url [1]{\endgroup\@href {#1}{\urlprefix }}%
\providecommand \urlprefix  [0]{URL }%
\providecommand \Eprint [0]{\href }%
\providecommand \doibase [0]{https://doi.org/}%
\providecommand \selectlanguage [0]{\@gobble}%
\providecommand \bibinfo  [0]{\@secondoftwo}%
\providecommand \bibfield  [0]{\@secondoftwo}%
\providecommand \translation [1]{[#1]}%
\providecommand \BibitemOpen [0]{}%
\providecommand \bibitemStop [0]{}%
\providecommand \bibitemNoStop [0]{.\EOS\space}%
\providecommand \EOS [0]{\spacefactor3000\relax}%
\providecommand \BibitemShut  [1]{\csname bibitem#1\endcsname}%
\let\auto@bib@innerbib\@empty
\bibitem [{\citenamefont {Dirac}(1938)}]{Dirac:1938nz}%
  \BibitemOpen
  \bibfield  {author} {\bibinfo {author} {\bibfnamefont {P.~A.~M.}\
  \bibnamefont {Dirac}},\ }\bibfield  {title} {\bibinfo {title} {{Classical
  theory of radiating electrons}},\ }\href
  {https://doi.org/10.1098/rspa.1938.0124} {\bibfield  {journal} {\bibinfo
  {journal} {Proc. Roy. Soc. Lond. A}\ }\textbf {\bibinfo {volume} {167}},\
  \bibinfo {pages} {148} (\bibinfo {year} {1938})}\BibitemShut {NoStop}%
\bibitem [{\citenamefont {DeWitt}\ and\ \citenamefont
  {Brehme}(1960)}]{DeWitt:1960fc}%
  \BibitemOpen
  \bibfield  {author} {\bibinfo {author} {\bibfnamefont {B.~S.}\ \bibnamefont
  {DeWitt}}\ and\ \bibinfo {author} {\bibfnamefont {R.~W.}\ \bibnamefont
  {Brehme}},\ }\bibfield  {title} {\bibinfo {title} {{Radiation damping in a
  gravitational field}},\ }\href {https://doi.org/10.1016/0003-4916(60)90030-0}
  {\bibfield  {journal} {\bibinfo  {journal} {Annals Phys.}\ }\textbf {\bibinfo
  {volume} {9}},\ \bibinfo {pages} {220} (\bibinfo {year} {1960})}\BibitemShut
  {NoStop}%
\bibitem [{\citenamefont {Hobbs}(1968)}]{hobbs1968vierbein}%
  \BibitemOpen
  \bibfield  {author} {\bibinfo {author} {\bibfnamefont {J.}~\bibnamefont
  {Hobbs}},\ }\bibfield  {title} {\bibinfo {title} {A vierbein formalism of
  radiation damping},\ }\href@noop {} {\bibfield  {journal} {\bibinfo
  {journal} {Annals of Physics}\ }\textbf {\bibinfo {volume} {47}},\ \bibinfo
  {pages} {141} (\bibinfo {year} {1968})}\BibitemShut {NoStop}%
\bibitem [{\citenamefont {Quinn}\ and\ \citenamefont
  {Wald}(1997)}]{Quinn:1996am}%
  \BibitemOpen
  \bibfield  {author} {\bibinfo {author} {\bibfnamefont {T.~C.}\ \bibnamefont
  {Quinn}}\ and\ \bibinfo {author} {\bibfnamefont {R.~M.}\ \bibnamefont
  {Wald}},\ }\bibfield  {title} {\bibinfo {title} {{An Axiomatic approach to
  electromagnetic and gravitational radiation reaction of particles in curved
  space-time}},\ }\href {https://doi.org/10.1103/PhysRevD.56.3381} {\bibfield
  {journal} {\bibinfo  {journal} {Phys. Rev. D}\ }\textbf {\bibinfo {volume}
  {56}},\ \bibinfo {pages} {3381} (\bibinfo {year} {1997})},\ \Eprint
  {https://arxiv.org/abs/gr-qc/9610053} {arXiv:gr-qc/9610053} \BibitemShut
  {NoStop}%
\bibitem [{\citenamefont {Detweiler}\ and\ \citenamefont
  {Whiting}(2003)}]{Detweiler:2002mi}%
  \BibitemOpen
  \bibfield  {author} {\bibinfo {author} {\bibfnamefont {S.~L.}\ \bibnamefont
  {Detweiler}}\ and\ \bibinfo {author} {\bibfnamefont {B.~F.}\ \bibnamefont
  {Whiting}},\ }\bibfield  {title} {\bibinfo {title} {{Selfforce via a Green's
  function decomposition}},\ }\href
  {https://doi.org/10.1103/PhysRevD.67.024025} {\bibfield  {journal} {\bibinfo
  {journal} {Phys. Rev. D}\ }\textbf {\bibinfo {volume} {67}},\ \bibinfo
  {pages} {024025} (\bibinfo {year} {2003})},\ \Eprint
  {https://arxiv.org/abs/gr-qc/0202086} {arXiv:gr-qc/0202086} \BibitemShut
  {NoStop}%
\bibitem [{\citenamefont {Barack}\ and\ \citenamefont
  {Ori}(2000)}]{Barack:1999wf}%
  \BibitemOpen
  \bibfield  {author} {\bibinfo {author} {\bibfnamefont {L.}~\bibnamefont
  {Barack}}\ and\ \bibinfo {author} {\bibfnamefont {A.}~\bibnamefont {Ori}},\
  }\bibfield  {title} {\bibinfo {title} {{Mode sum regularization approach for
  the selfforce in black hole space-time}},\ }\href
  {https://doi.org/10.1103/PhysRevD.61.061502} {\bibfield  {journal} {\bibinfo
  {journal} {Phys. Rev. D}\ }\textbf {\bibinfo {volume} {61}},\ \bibinfo
  {pages} {061502} (\bibinfo {year} {2000})},\ \Eprint
  {https://arxiv.org/abs/gr-qc/9912010} {arXiv:gr-qc/9912010} \BibitemShut
  {NoStop}%
\bibitem [{\citenamefont {Barack}(2001)}]{Barack:2001bw}%
  \BibitemOpen
  \bibfield  {author} {\bibinfo {author} {\bibfnamefont {L.}~\bibnamefont
  {Barack}},\ }\bibfield  {title} {\bibinfo {title} {{Gravitational selfforce
  by mode sum regularization}},\ }\href
  {https://doi.org/10.1103/PhysRevD.64.084021} {\bibfield  {journal} {\bibinfo
  {journal} {Phys. Rev. D}\ }\textbf {\bibinfo {volume} {64}},\ \bibinfo
  {pages} {084021} (\bibinfo {year} {2001})},\ \Eprint
  {https://arxiv.org/abs/gr-qc/0105040} {arXiv:gr-qc/0105040} \BibitemShut
  {NoStop}%
\bibitem [{\citenamefont {Haas}(2011)}]{Haas:2011np}%
  \BibitemOpen
  \bibfield  {author} {\bibinfo {author} {\bibfnamefont {R.}~\bibnamefont
  {Haas}},\ }\bibfield  {title} {\bibinfo {title} {{Time domain calculation of
  the electromagnetic self-force on eccentric geodesics in Schwarzschild
  spacetime}},\ }\href@noop {} {\  (\bibinfo {year} {2011})},\ \Eprint
  {https://arxiv.org/abs/1112.3707} {arXiv:1112.3707 [gr-qc]} \BibitemShut
  {NoStop}%
\bibitem [{\citenamefont {Torres}\ and\ \citenamefont
  {Dolan}(2022)}]{Torres:2020fye}%
  \BibitemOpen
  \bibfield  {author} {\bibinfo {author} {\bibfnamefont {T.}~\bibnamefont
  {Torres}}\ and\ \bibinfo {author} {\bibfnamefont {S.~R.}\ \bibnamefont
  {Dolan}},\ }\bibfield  {title} {\bibinfo {title} {{Electromagnetic self-force
  on a charged particle on Kerr spacetime: Equatorial circular orbits}},\
  }\href {https://doi.org/10.1103/PhysRevD.106.024024} {\bibfield  {journal}
  {\bibinfo  {journal} {Phys. Rev. D}\ }\textbf {\bibinfo {volume} {106}},\
  \bibinfo {pages} {024024} (\bibinfo {year} {2022})},\ \Eprint
  {https://arxiv.org/abs/2008.12703} {arXiv:2008.12703 [gr-qc]} \BibitemShut
  {NoStop}%
\bibitem [{\citenamefont {Detweiler}\ \emph {et~al.}(2003)\citenamefont
  {Detweiler}, \citenamefont {Messaritaki},\ and\ \citenamefont
  {Whiting}}]{Detweiler:2002gi}%
  \BibitemOpen
  \bibfield  {author} {\bibinfo {author} {\bibfnamefont {S.~L.}\ \bibnamefont
  {Detweiler}}, \bibinfo {author} {\bibfnamefont {E.}~\bibnamefont
  {Messaritaki}},\ and\ \bibinfo {author} {\bibfnamefont {B.~F.}\ \bibnamefont
  {Whiting}},\ }\bibfield  {title} {\bibinfo {title} {{Selfforce of a scalar
  field for circular orbits about a Schwarzschild black hole}},\ }\href
  {https://doi.org/10.1103/PhysRevD.67.104016} {\bibfield  {journal} {\bibinfo
  {journal} {Phys. Rev. D}\ }\textbf {\bibinfo {volume} {67}},\ \bibinfo
  {pages} {104016} (\bibinfo {year} {2003})},\ \Eprint
  {https://arxiv.org/abs/gr-qc/0205079} {arXiv:gr-qc/0205079} \BibitemShut
  {NoStop}%
\bibitem [{\citenamefont {Barack}\ and\ \citenamefont
  {Sago}(2007)}]{Barack:2007tm}%
  \BibitemOpen
  \bibfield  {author} {\bibinfo {author} {\bibfnamefont {L.}~\bibnamefont
  {Barack}}\ and\ \bibinfo {author} {\bibfnamefont {N.}~\bibnamefont {Sago}},\
  }\bibfield  {title} {\bibinfo {title} {{Gravitational self force on a
  particle in circular orbit around a Schwarzschild black hole}},\ }\href
  {https://doi.org/10.1103/PhysRevD.75.064021} {\bibfield  {journal} {\bibinfo
  {journal} {Phys. Rev. D}\ }\textbf {\bibinfo {volume} {75}},\ \bibinfo
  {pages} {064021} (\bibinfo {year} {2007})},\ \Eprint
  {https://arxiv.org/abs/gr-qc/0701069} {arXiv:gr-qc/0701069} \BibitemShut
  {NoStop}%
\bibitem [{\citenamefont {Detweiler}(2008)}]{Detweiler:2008ft}%
  \BibitemOpen
  \bibfield  {author} {\bibinfo {author} {\bibfnamefont {S.~L.}\ \bibnamefont
  {Detweiler}},\ }\bibfield  {title} {\bibinfo {title} {{A Consequence of the
  gravitational self-force for circular orbits of the Schwarzschild
  geometry}},\ }\href {https://doi.org/10.1103/PhysRevD.77.124026} {\bibfield
  {journal} {\bibinfo  {journal} {Phys. Rev. D}\ }\textbf {\bibinfo {volume}
  {77}},\ \bibinfo {pages} {124026} (\bibinfo {year} {2008})},\ \Eprint
  {https://arxiv.org/abs/0804.3529} {arXiv:0804.3529 [gr-qc]} \BibitemShut
  {NoStop}%
\bibitem [{\citenamefont {Akcay}(2011)}]{Akcay:2010dx}%
  \BibitemOpen
  \bibfield  {author} {\bibinfo {author} {\bibfnamefont {S.}~\bibnamefont
  {Akcay}},\ }\bibfield  {title} {\bibinfo {title} {{A Fast Frequency-Domain
  Algorithm for Gravitational Self-Force: I. Circular Orbits in Schwarzschild
  Spacetime}},\ }\href {https://doi.org/10.1103/PhysRevD.83.124026} {\bibfield
  {journal} {\bibinfo  {journal} {Phys. Rev. D}\ }\textbf {\bibinfo {volume}
  {83}},\ \bibinfo {pages} {124026} (\bibinfo {year} {2011})},\ \Eprint
  {https://arxiv.org/abs/1012.5860} {arXiv:1012.5860 [gr-qc]} \BibitemShut
  {NoStop}%
\bibitem [{\citenamefont {Barack}\ and\ \citenamefont
  {Sago}(2010)}]{Barack:2010tm}%
  \BibitemOpen
  \bibfield  {author} {\bibinfo {author} {\bibfnamefont {L.}~\bibnamefont
  {Barack}}\ and\ \bibinfo {author} {\bibfnamefont {N.}~\bibnamefont {Sago}},\
  }\bibfield  {title} {\bibinfo {title} {{Gravitational self-force on a
  particle in eccentric orbit around a Schwarzschild black hole}},\ }\href
  {https://doi.org/10.1103/PhysRevD.81.084021} {\bibfield  {journal} {\bibinfo
  {journal} {Phys. Rev. D}\ }\textbf {\bibinfo {volume} {81}},\ \bibinfo
  {pages} {084021} (\bibinfo {year} {2010})},\ \Eprint
  {https://arxiv.org/abs/1002.2386} {arXiv:1002.2386 [gr-qc]} \BibitemShut
  {NoStop}%
\bibitem [{\citenamefont {Akcay}\ \emph {et~al.}(2013)\citenamefont {Akcay},
  \citenamefont {Warburton},\ and\ \citenamefont {Barack}}]{Akcay:2013wfa}%
  \BibitemOpen
  \bibfield  {author} {\bibinfo {author} {\bibfnamefont {S.}~\bibnamefont
  {Akcay}}, \bibinfo {author} {\bibfnamefont {N.}~\bibnamefont {Warburton}},\
  and\ \bibinfo {author} {\bibfnamefont {L.}~\bibnamefont {Barack}},\
  }\bibfield  {title} {\bibinfo {title} {{Frequency-domain algorithm for the
  Lorenz-gauge gravitational self-force}},\ }\href
  {https://doi.org/10.1103/PhysRevD.88.104009} {\bibfield  {journal} {\bibinfo
  {journal} {Phys. Rev. D}\ }\textbf {\bibinfo {volume} {88}},\ \bibinfo
  {pages} {104009} (\bibinfo {year} {2013})},\ \Eprint
  {https://arxiv.org/abs/1308.5223} {arXiv:1308.5223 [gr-qc]} \BibitemShut
  {NoStop}%
\bibitem [{\citenamefont {Osburn}\ \emph {et~al.}(2014)\citenamefont {Osburn},
  \citenamefont {Forseth}, \citenamefont {Evans},\ and\ \citenamefont
  {Hopper}}]{Osburn:2014hoa}%
  \BibitemOpen
  \bibfield  {author} {\bibinfo {author} {\bibfnamefont {T.}~\bibnamefont
  {Osburn}}, \bibinfo {author} {\bibfnamefont {E.}~\bibnamefont {Forseth}},
  \bibinfo {author} {\bibfnamefont {C.~R.}\ \bibnamefont {Evans}},\ and\
  \bibinfo {author} {\bibfnamefont {S.}~\bibnamefont {Hopper}},\ }\bibfield
  {title} {\bibinfo {title} {{Lorenz gauge gravitational self-force
  calculations of eccentric binaries using a frequency domain procedure}},\
  }\href {https://doi.org/10.1103/PhysRevD.90.104031} {\bibfield  {journal}
  {\bibinfo  {journal} {Phys. Rev. D}\ }\textbf {\bibinfo {volume} {90}},\
  \bibinfo {pages} {104031} (\bibinfo {year} {2014})},\ \Eprint
  {https://arxiv.org/abs/1409.4419} {arXiv:1409.4419 [gr-qc]} \BibitemShut
  {NoStop}%
\bibitem [{\citenamefont {van~de Meent}(2018)}]{vandeMeent:2017bcc}%
  \BibitemOpen
  \bibfield  {author} {\bibinfo {author} {\bibfnamefont {M.}~\bibnamefont
  {van~de Meent}},\ }\bibfield  {title} {\bibinfo {title} {{Gravitational
  self-force on generic bound geodesics in Kerr spacetime}},\ }\href
  {https://doi.org/10.1103/PhysRevD.97.104033} {\bibfield  {journal} {\bibinfo
  {journal} {Phys. Rev. D}\ }\textbf {\bibinfo {volume} {97}},\ \bibinfo
  {pages} {104033} (\bibinfo {year} {2018})},\ \Eprint
  {https://arxiv.org/abs/1711.09607} {arXiv:1711.09607 [gr-qc]} \BibitemShut
  {NoStop}%
\bibitem [{\citenamefont {Pound}(2010)}]{Pound:2009sm}%
  \BibitemOpen
  \bibfield  {author} {\bibinfo {author} {\bibfnamefont {A.}~\bibnamefont
  {Pound}},\ }\bibfield  {title} {\bibinfo {title} {{Self-consistent
  gravitational self-force}},\ }\href
  {https://doi.org/10.1103/PhysRevD.81.024023} {\bibfield  {journal} {\bibinfo
  {journal} {Phys. Rev. D}\ }\textbf {\bibinfo {volume} {81}},\ \bibinfo
  {pages} {024023} (\bibinfo {year} {2010})},\ \Eprint
  {https://arxiv.org/abs/0907.5197} {arXiv:0907.5197 [gr-qc]} \BibitemShut
  {NoStop}%
\bibitem [{\citenamefont {Warburton}\ \emph {et~al.}(2012)\citenamefont
  {Warburton}, \citenamefont {Akcay}, \citenamefont {Barack}, \citenamefont
  {Gair},\ and\ \citenamefont {Sago}}]{Warburton:2011fk}%
  \BibitemOpen
  \bibfield  {author} {\bibinfo {author} {\bibfnamefont {N.}~\bibnamefont
  {Warburton}}, \bibinfo {author} {\bibfnamefont {S.}~\bibnamefont {Akcay}},
  \bibinfo {author} {\bibfnamefont {L.}~\bibnamefont {Barack}}, \bibinfo
  {author} {\bibfnamefont {J.~R.}\ \bibnamefont {Gair}},\ and\ \bibinfo
  {author} {\bibfnamefont {N.}~\bibnamefont {Sago}},\ }\bibfield  {title}
  {\bibinfo {title} {{Evolution of inspiral orbits around a Schwarzschild black
  hole}},\ }\href {https://doi.org/10.1103/PhysRevD.85.061501} {\bibfield
  {journal} {\bibinfo  {journal} {Phys. Rev. D}\ }\textbf {\bibinfo {volume}
  {85}},\ \bibinfo {pages} {061501} (\bibinfo {year} {2012})},\ \Eprint
  {https://arxiv.org/abs/1111.6908} {arXiv:1111.6908 [gr-qc]} \BibitemShut
  {NoStop}%
\bibitem [{\citenamefont {Osburn}\ \emph {et~al.}(2016)\citenamefont {Osburn},
  \citenamefont {Warburton},\ and\ \citenamefont {Evans}}]{Osburn:2015duj}%
  \BibitemOpen
  \bibfield  {author} {\bibinfo {author} {\bibfnamefont {T.}~\bibnamefont
  {Osburn}}, \bibinfo {author} {\bibfnamefont {N.}~\bibnamefont {Warburton}},\
  and\ \bibinfo {author} {\bibfnamefont {C.~R.}\ \bibnamefont {Evans}},\
  }\bibfield  {title} {\bibinfo {title} {{Highly eccentric inspirals into a
  black hole}},\ }\href {https://doi.org/10.1103/PhysRevD.93.064024} {\bibfield
   {journal} {\bibinfo  {journal} {Phys. Rev. D}\ }\textbf {\bibinfo {volume}
  {93}},\ \bibinfo {pages} {064024} (\bibinfo {year} {2016})},\ \Eprint
  {https://arxiv.org/abs/1511.01498} {arXiv:1511.01498 [gr-qc]} \BibitemShut
  {NoStop}%
\bibitem [{\citenamefont {Warburton}\ \emph {et~al.}(2017)\citenamefont
  {Warburton}, \citenamefont {Osburn},\ and\ \citenamefont
  {Evans}}]{Warburton:2017sxk}%
  \BibitemOpen
  \bibfield  {author} {\bibinfo {author} {\bibfnamefont {N.}~\bibnamefont
  {Warburton}}, \bibinfo {author} {\bibfnamefont {T.}~\bibnamefont {Osburn}},\
  and\ \bibinfo {author} {\bibfnamefont {C.~R.}\ \bibnamefont {Evans}},\
  }\bibfield  {title} {\bibinfo {title} {{Evolution of small-mass-ratio
  binaries with a spinning secondary}},\ }\href
  {https://doi.org/10.1103/PhysRevD.96.084057} {\bibfield  {journal} {\bibinfo
  {journal} {Phys. Rev. D}\ }\textbf {\bibinfo {volume} {96}},\ \bibinfo
  {pages} {084057} (\bibinfo {year} {2017})},\ \Eprint
  {https://arxiv.org/abs/1708.03720} {arXiv:1708.03720 [gr-qc]} \BibitemShut
  {NoStop}%
\bibitem [{\citenamefont {Van De~Meent}\ and\ \citenamefont
  {Warburton}(2018)}]{VanDeMeent:2018cgn}%
  \BibitemOpen
  \bibfield  {author} {\bibinfo {author} {\bibfnamefont {M.}~\bibnamefont {Van
  De~Meent}}\ and\ \bibinfo {author} {\bibfnamefont {N.}~\bibnamefont
  {Warburton}},\ }\bibfield  {title} {\bibinfo {title} {{Fast Self-forced
  Inspirals}},\ }\href {https://doi.org/10.1088/1361-6382/aac8ce} {\bibfield
  {journal} {\bibinfo  {journal} {Class. Quant. Grav.}\ }\textbf {\bibinfo
  {volume} {35}},\ \bibinfo {pages} {144003} (\bibinfo {year} {2018})},\
  \Eprint {https://arxiv.org/abs/1802.05281} {arXiv:1802.05281 [gr-qc]}
  \BibitemShut {NoStop}%
\bibitem [{\citenamefont {Lynch}\ \emph {et~al.}(2022)\citenamefont {Lynch},
  \citenamefont {van~de Meent},\ and\ \citenamefont
  {Warburton}}]{Lynch:2021ogr}%
  \BibitemOpen
  \bibfield  {author} {\bibinfo {author} {\bibfnamefont {P.}~\bibnamefont
  {Lynch}}, \bibinfo {author} {\bibfnamefont {M.}~\bibnamefont {van~de
  Meent}},\ and\ \bibinfo {author} {\bibfnamefont {N.}~\bibnamefont
  {Warburton}},\ }\bibfield  {title} {\bibinfo {title} {{Eccentric self-forced
  inspirals into a rotating black hole}},\ }\href
  {https://doi.org/10.1088/1361-6382/ac7507} {\bibfield  {journal} {\bibinfo
  {journal} {Class. Quant. Grav.}\ }\textbf {\bibinfo {volume} {39}},\ \bibinfo
  {pages} {145004} (\bibinfo {year} {2022})},\ \Eprint
  {https://arxiv.org/abs/2112.05651} {arXiv:2112.05651 [gr-qc]} \BibitemShut
  {NoStop}%
\bibitem [{\citenamefont {Lynch}\ \emph {et~al.}(2024)\citenamefont {Lynch},
  \citenamefont {van~de Meent},\ and\ \citenamefont
  {Warburton}}]{Lynch:2023gpu}%
  \BibitemOpen
  \bibfield  {author} {\bibinfo {author} {\bibfnamefont {P.}~\bibnamefont
  {Lynch}}, \bibinfo {author} {\bibfnamefont {M.}~\bibnamefont {van~de
  Meent}},\ and\ \bibinfo {author} {\bibfnamefont {N.}~\bibnamefont
  {Warburton}},\ }\bibfield  {title} {\bibinfo {title} {{Self-forced inspirals
  with spin-orbit precession}},\ }\href
  {https://doi.org/10.1103/PhysRevD.109.084072} {\bibfield  {journal} {\bibinfo
   {journal} {Phys. Rev. D}\ }\textbf {\bibinfo {volume} {109}},\ \bibinfo
  {pages} {084072} (\bibinfo {year} {2024})},\ \Eprint
  {https://arxiv.org/abs/2305.10533} {arXiv:2305.10533 [gr-qc]} \BibitemShut
  {NoStop}%
\bibitem [{\citenamefont {Zhang}\ and\ \citenamefont
  {Gong}(2026)}]{Zhang:2026zqs}%
  \BibitemOpen
  \bibfield  {author} {\bibinfo {author} {\bibfnamefont {B.}~\bibnamefont
  {Zhang}}\ and\ \bibinfo {author} {\bibfnamefont {Y.}~\bibnamefont {Gong}},\
  }\bibfield  {title} {\bibinfo {title} {{Self-Forces as Nonlocal Probes of
  Gravastar Interiors}},\ }\href@noop {} {\  (\bibinfo {year} {2026})},\
  \Eprint {https://arxiv.org/abs/2608.00121} {arXiv:2608.00121 [gr-qc]}
  \BibitemShut {NoStop}%
\bibitem [{\citenamefont {Rosenthal}(2006{\natexlab{a}})}]{Rosenthal:2006nh}%
  \BibitemOpen
  \bibfield  {author} {\bibinfo {author} {\bibfnamefont {E.}~\bibnamefont
  {Rosenthal}},\ }\bibfield  {title} {\bibinfo {title} {{Construction of the
  second-order gravitational perturbations produced by a compact object}},\
  }\href {https://doi.org/10.1103/PhysRevD.73.044034} {\bibfield  {journal}
  {\bibinfo  {journal} {Phys. Rev. D}\ }\textbf {\bibinfo {volume} {73}},\
  \bibinfo {pages} {044034} (\bibinfo {year} {2006}{\natexlab{a}})},\ \Eprint
  {https://arxiv.org/abs/gr-qc/0602066} {arXiv:gr-qc/0602066} \BibitemShut
  {NoStop}%
\bibitem [{\citenamefont {Rosenthal}(2006{\natexlab{b}})}]{Rosenthal:2006iy}%
  \BibitemOpen
  \bibfield  {author} {\bibinfo {author} {\bibfnamefont {E.}~\bibnamefont
  {Rosenthal}},\ }\bibfield  {title} {\bibinfo {title} {{Second-order
  gravitational self-force}},\ }\href
  {https://doi.org/10.1103/PhysRevD.74.084018} {\bibfield  {journal} {\bibinfo
  {journal} {Phys. Rev. D}\ }\textbf {\bibinfo {volume} {74}},\ \bibinfo
  {pages} {084018} (\bibinfo {year} {2006}{\natexlab{b}})},\ \Eprint
  {https://arxiv.org/abs/gr-qc/0609069} {arXiv:gr-qc/0609069} \BibitemShut
  {NoStop}%
\bibitem [{\citenamefont {Pound}(2012)}]{Pound:2012nt}%
  \BibitemOpen
  \bibfield  {author} {\bibinfo {author} {\bibfnamefont {A.}~\bibnamefont
  {Pound}},\ }\bibfield  {title} {\bibinfo {title} {{Second-order gravitational
  self-force}},\ }\href {https://doi.org/10.1103/PhysRevLett.109.051101}
  {\bibfield  {journal} {\bibinfo  {journal} {Phys. Rev. Lett.}\ }\textbf
  {\bibinfo {volume} {109}},\ \bibinfo {pages} {051101} (\bibinfo {year}
  {2012})},\ \Eprint {https://arxiv.org/abs/1201.5089} {arXiv:1201.5089
  [gr-qc]} \BibitemShut {NoStop}%
\bibitem [{\citenamefont {Gralla}(2012)}]{Gralla:2012db}%
  \BibitemOpen
  \bibfield  {author} {\bibinfo {author} {\bibfnamefont {S.~E.}\ \bibnamefont
  {Gralla}},\ }\bibfield  {title} {\bibinfo {title} {{Second Order
  Gravitational Self Force}},\ }\href
  {https://doi.org/10.1103/PhysRevD.85.124011} {\bibfield  {journal} {\bibinfo
  {journal} {Phys. Rev. D}\ }\textbf {\bibinfo {volume} {85}},\ \bibinfo
  {pages} {124011} (\bibinfo {year} {2012})},\ \Eprint
  {https://arxiv.org/abs/1203.3189} {arXiv:1203.3189 [gr-qc]} \BibitemShut
  {NoStop}%
\bibitem [{\citenamefont {Pound}(2017)}]{Pound:2017psq}%
  \BibitemOpen
  \bibfield  {author} {\bibinfo {author} {\bibfnamefont {A.}~\bibnamefont
  {Pound}},\ }\bibfield  {title} {\bibinfo {title} {{Nonlinear gravitational
  self-force: second-order equation of motion}},\ }\href
  {https://doi.org/10.1103/PhysRevD.95.104056} {\bibfield  {journal} {\bibinfo
  {journal} {Phys. Rev. D}\ }\textbf {\bibinfo {volume} {95}},\ \bibinfo
  {pages} {104056} (\bibinfo {year} {2017})},\ \Eprint
  {https://arxiv.org/abs/1703.02836} {arXiv:1703.02836 [gr-qc]} \BibitemShut
  {NoStop}%
\bibitem [{\citenamefont {Pound}\ \emph {et~al.}(2020)\citenamefont {Pound},
  \citenamefont {Wardell}, \citenamefont {Warburton},\ and\ \citenamefont
  {Miller}}]{Pound:2019lzj}%
  \BibitemOpen
  \bibfield  {author} {\bibinfo {author} {\bibfnamefont {A.}~\bibnamefont
  {Pound}}, \bibinfo {author} {\bibfnamefont {B.}~\bibnamefont {Wardell}},
  \bibinfo {author} {\bibfnamefont {N.}~\bibnamefont {Warburton}},\ and\
  \bibinfo {author} {\bibfnamefont {J.}~\bibnamefont {Miller}},\ }\bibfield
  {title} {\bibinfo {title} {{Second-Order Self-Force Calculation of
  Gravitational Binding Energy in Compact Binaries}},\ }\href
  {https://doi.org/10.1103/PhysRevLett.124.021101} {\bibfield  {journal}
  {\bibinfo  {journal} {Phys. Rev. Lett.}\ }\textbf {\bibinfo {volume} {124}},\
  \bibinfo {pages} {021101} (\bibinfo {year} {2020})},\ \Eprint
  {https://arxiv.org/abs/1908.07419} {arXiv:1908.07419 [gr-qc]} \BibitemShut
  {NoStop}%
\bibitem [{\citenamefont {Miller}\ and\ \citenamefont
  {Pound}(2021)}]{Miller:2020bft}%
  \BibitemOpen
  \bibfield  {author} {\bibinfo {author} {\bibfnamefont {J.}~\bibnamefont
  {Miller}}\ and\ \bibinfo {author} {\bibfnamefont {A.}~\bibnamefont {Pound}},\
  }\bibfield  {title} {\bibinfo {title} {{Two-timescale evolution of
  extreme-mass-ratio inspirals: waveform generation scheme for quasicircular
  orbits in Schwarzschild spacetime}},\ }\href
  {https://doi.org/10.1103/PhysRevD.103.064048} {\bibfield  {journal} {\bibinfo
   {journal} {Phys. Rev. D}\ }\textbf {\bibinfo {volume} {103}},\ \bibinfo
  {pages} {064048} (\bibinfo {year} {2021})},\ \Eprint
  {https://arxiv.org/abs/2006.11263} {arXiv:2006.11263 [gr-qc]} \BibitemShut
  {NoStop}%
\bibitem [{\citenamefont {Wardell}\ \emph {et~al.}(2023)\citenamefont
  {Wardell}, \citenamefont {Pound}, \citenamefont {Warburton}, \citenamefont
  {Miller}, \citenamefont {Durkan},\ and\ \citenamefont
  {Le~Tiec}}]{Wardell:2021fyy}%
  \BibitemOpen
  \bibfield  {author} {\bibinfo {author} {\bibfnamefont {B.}~\bibnamefont
  {Wardell}}, \bibinfo {author} {\bibfnamefont {A.}~\bibnamefont {Pound}},
  \bibinfo {author} {\bibfnamefont {N.}~\bibnamefont {Warburton}}, \bibinfo
  {author} {\bibfnamefont {J.}~\bibnamefont {Miller}}, \bibinfo {author}
  {\bibfnamefont {L.}~\bibnamefont {Durkan}},\ and\ \bibinfo {author}
  {\bibfnamefont {A.}~\bibnamefont {Le~Tiec}},\ }\bibfield  {title} {\bibinfo
  {title} {{Gravitational Waveforms for Compact Binaries from Second-Order
  Self-Force Theory}},\ }\href {https://doi.org/10.1103/PhysRevLett.130.241402}
  {\bibfield  {journal} {\bibinfo  {journal} {Phys. Rev. Lett.}\ }\textbf
  {\bibinfo {volume} {130}},\ \bibinfo {pages} {241402} (\bibinfo {year}
  {2023})},\ \Eprint {https://arxiv.org/abs/2112.12265} {arXiv:2112.12265
  [gr-qc]} \BibitemShut {NoStop}%
\bibitem [{\citenamefont {Upton}\ and\ \citenamefont
  {Pound}(2021)}]{Upton:2021oxf}%
  \BibitemOpen
  \bibfield  {author} {\bibinfo {author} {\bibfnamefont {S.~D.}\ \bibnamefont
  {Upton}}\ and\ \bibinfo {author} {\bibfnamefont {A.}~\bibnamefont {Pound}},\
  }\bibfield  {title} {\bibinfo {title} {{Second-order gravitational self-force
  in a highly regular gauge}},\ }\href
  {https://doi.org/10.1103/PhysRevD.103.124016} {\bibfield  {journal} {\bibinfo
   {journal} {Phys. Rev. D}\ }\textbf {\bibinfo {volume} {103}},\ \bibinfo
  {pages} {124016} (\bibinfo {year} {2021})},\ \Eprint
  {https://arxiv.org/abs/2101.11409} {arXiv:2101.11409 [gr-qc]} \BibitemShut
  {NoStop}%
\bibitem [{\citenamefont {Upton}(2024)}]{Upton:2023tcv}%
  \BibitemOpen
  \bibfield  {author} {\bibinfo {author} {\bibfnamefont {S.~D.}\ \bibnamefont
  {Upton}},\ }\bibfield  {title} {\bibinfo {title} {{Second-order gravitational
  self-force in a highly regular gauge: Covariant and coordinate punctures}},\
  }\href {https://doi.org/10.1103/PhysRevD.109.044021} {\bibfield  {journal}
  {\bibinfo  {journal} {Phys. Rev. D}\ }\textbf {\bibinfo {volume} {109}},\
  \bibinfo {pages} {044021} (\bibinfo {year} {2024})},\ \Eprint
  {https://arxiv.org/abs/2309.03778} {arXiv:2309.03778 [gr-qc]} \BibitemShut
  {NoStop}%
\bibitem [{\citenamefont {Pound}\ and\ \citenamefont
  {Miller}(2014)}]{Pound:2014xva}%
  \BibitemOpen
  \bibfield  {author} {\bibinfo {author} {\bibfnamefont {A.}~\bibnamefont
  {Pound}}\ and\ \bibinfo {author} {\bibfnamefont {J.}~\bibnamefont {Miller}},\
  }\bibfield  {title} {\bibinfo {title} {{Practical, covariant puncture for
  second-order self-force calculations}},\ }\href
  {https://doi.org/10.1103/PhysRevD.89.104020} {\bibfield  {journal} {\bibinfo
  {journal} {Phys. Rev. D}\ }\textbf {\bibinfo {volume} {89}},\ \bibinfo
  {pages} {104020} (\bibinfo {year} {2014})},\ \Eprint
  {https://arxiv.org/abs/1403.1843} {arXiv:1403.1843 [gr-qc]} \BibitemShut
  {NoStop}%
\bibitem [{\citenamefont {Wardell}\ and\ \citenamefont
  {Warburton}(2015)}]{Wardell:2015ada}%
  \BibitemOpen
  \bibfield  {author} {\bibinfo {author} {\bibfnamefont {B.}~\bibnamefont
  {Wardell}}\ and\ \bibinfo {author} {\bibfnamefont {N.}~\bibnamefont
  {Warburton}},\ }\bibfield  {title} {\bibinfo {title} {{Applying the
  effective-source approach to frequency-domain self-force calculations:
  Lorenz-gauge gravitational perturbations}},\ }\href
  {https://doi.org/10.1103/PhysRevD.92.084019} {\bibfield  {journal} {\bibinfo
  {journal} {Phys. Rev. D}\ }\textbf {\bibinfo {volume} {92}},\ \bibinfo
  {pages} {084019} (\bibinfo {year} {2015})},\ \Eprint
  {https://arxiv.org/abs/1505.07841} {arXiv:1505.07841 [gr-qc]} \BibitemShut
  {NoStop}%
\bibitem [{\citenamefont {Barack}\ and\ \citenamefont
  {Golbourn}(2007)}]{Barack:2007jh}%
  \BibitemOpen
  \bibfield  {author} {\bibinfo {author} {\bibfnamefont {L.}~\bibnamefont
  {Barack}}\ and\ \bibinfo {author} {\bibfnamefont {D.~A.}\ \bibnamefont
  {Golbourn}},\ }\bibfield  {title} {\bibinfo {title} {{Scalar-field
  perturbations from a particle orbiting a black hole using numerical evolution
  in 2+1 dimensions}},\ }\href {https://doi.org/10.1103/PhysRevD.76.044020}
  {\bibfield  {journal} {\bibinfo  {journal} {Phys. Rev. D}\ }\textbf {\bibinfo
  {volume} {76}},\ \bibinfo {pages} {044020} (\bibinfo {year} {2007})},\
  \Eprint {https://arxiv.org/abs/0705.3620} {arXiv:0705.3620 [gr-qc]}
  \BibitemShut {NoStop}%
\bibitem [{\citenamefont {Vega}\ and\ \citenamefont
  {Detweiler}(2008)}]{Vega:2007mc}%
  \BibitemOpen
  \bibfield  {author} {\bibinfo {author} {\bibfnamefont {I.}~\bibnamefont
  {Vega}}\ and\ \bibinfo {author} {\bibfnamefont {S.~L.}\ \bibnamefont
  {Detweiler}},\ }\bibfield  {title} {\bibinfo {title} {{Regularization of
  fields for self-force problems in curved spacetime: Foundations and a
  time-domain application}},\ }\href
  {https://doi.org/10.1103/PhysRevD.77.084008} {\bibfield  {journal} {\bibinfo
  {journal} {Phys. Rev. D}\ }\textbf {\bibinfo {volume} {77}},\ \bibinfo
  {pages} {084008} (\bibinfo {year} {2008})},\ \Eprint
  {https://arxiv.org/abs/0712.4405} {arXiv:0712.4405 [gr-qc]} \BibitemShut
  {NoStop}%
\bibitem [{\citenamefont {Warburton}\ and\ \citenamefont
  {Wardell}(2014)}]{Warburton:2013lea}%
  \BibitemOpen
  \bibfield  {author} {\bibinfo {author} {\bibfnamefont {N.}~\bibnamefont
  {Warburton}}\ and\ \bibinfo {author} {\bibfnamefont {B.}~\bibnamefont
  {Wardell}},\ }\bibfield  {title} {\bibinfo {title} {{Applying the
  effective-source approach to frequency-domain self-force calculations}},\
  }\href {https://doi.org/10.1103/PhysRevD.89.044046} {\bibfield  {journal}
  {\bibinfo  {journal} {Phys. Rev. D}\ }\textbf {\bibinfo {volume} {89}},\
  \bibinfo {pages} {044046} (\bibinfo {year} {2014})},\ \Eprint
  {https://arxiv.org/abs/1311.3104} {arXiv:1311.3104 [gr-qc]} \BibitemShut
  {NoStop}%
\bibitem [{\citenamefont {Diener}\ \emph {et~al.}(2012)\citenamefont {Diener},
  \citenamefont {Vega}, \citenamefont {Wardell},\ and\ \citenamefont
  {Detweiler}}]{Diener:2011cc}%
  \BibitemOpen
  \bibfield  {author} {\bibinfo {author} {\bibfnamefont {P.}~\bibnamefont
  {Diener}}, \bibinfo {author} {\bibfnamefont {I.}~\bibnamefont {Vega}},
  \bibinfo {author} {\bibfnamefont {B.}~\bibnamefont {Wardell}},\ and\ \bibinfo
  {author} {\bibfnamefont {S.}~\bibnamefont {Detweiler}},\ }\bibfield  {title}
  {\bibinfo {title} {{Self-consistent orbital evolution of a particle around a
  Schwarzschild black hole}},\ }\href
  {https://doi.org/10.1103/PhysRevLett.108.191102} {\bibfield  {journal}
  {\bibinfo  {journal} {Phys. Rev. Lett.}\ }\textbf {\bibinfo {volume} {108}},\
  \bibinfo {pages} {191102} (\bibinfo {year} {2012})},\ \Eprint
  {https://arxiv.org/abs/1112.4821} {arXiv:1112.4821 [gr-qc]} \BibitemShut
  {NoStop}%
\bibitem [{\citenamefont {Leather}\ and\ \citenamefont
  {Warburton}(2023)}]{Leather:2023dzj}%
  \BibitemOpen
  \bibfield  {author} {\bibinfo {author} {\bibfnamefont {B.}~\bibnamefont
  {Leather}}\ and\ \bibinfo {author} {\bibfnamefont {N.}~\bibnamefont
  {Warburton}},\ }\bibfield  {title} {\bibinfo {title} {{Applying the
  effective-source approach to frequency-domain self-force calculations for
  eccentric orbits}},\ }\href {https://doi.org/10.1103/PhysRevD.108.084045}
  {\bibfield  {journal} {\bibinfo  {journal} {Phys. Rev. D}\ }\textbf {\bibinfo
  {volume} {108}},\ \bibinfo {pages} {084045} (\bibinfo {year} {2023})},\
  \Eprint {https://arxiv.org/abs/2306.17221} {arXiv:2306.17221 [gr-qc]}
  \BibitemShut {NoStop}%
\bibitem [{\citenamefont {Zhang}\ \emph {et~al.}(2025)\citenamefont {Zhang},
  \citenamefont {Cai}, \citenamefont {Fu}, \citenamefont {Gong}, \citenamefont
  {Lu},\ and\ \citenamefont {Zhou}}]{Zhang:2025eqz}%
  \BibitemOpen
  \bibfield  {author} {\bibinfo {author} {\bibfnamefont {C.}~\bibnamefont
  {Zhang}}, \bibinfo {author} {\bibfnamefont {R.-g.}\ \bibnamefont {Cai}},
  \bibinfo {author} {\bibfnamefont {G.}~\bibnamefont {Fu}}, \bibinfo {author}
  {\bibfnamefont {Y.}~\bibnamefont {Gong}}, \bibinfo {author} {\bibfnamefont
  {X.}~\bibnamefont {Lu}},\ and\ \bibinfo {author} {\bibfnamefont
  {W.}~\bibnamefont {Zhou}},\ }\bibfield  {title} {\bibinfo {title} {{Generic
  effective sources for first-order in mass-ratio gravitational self-force
  calculations in Schwarzschild spacetime}},\ }\href
  {https://doi.org/10.1103/z9jr-qxc7} {\bibfield  {journal} {\bibinfo
  {journal} {Phys. Rev. D}\ }\textbf {\bibinfo {volume} {112}},\ \bibinfo
  {pages} {104069} (\bibinfo {year} {2025})},\ \Eprint
  {https://arxiv.org/abs/2505.19732} {arXiv:2505.19732 [gr-qc]} \BibitemShut
  {NoStop}%
\bibitem [{\citenamefont {Zhang}\ \emph {et~al.}(2026)\citenamefont {Zhang},
  \citenamefont {Gong}, \citenamefont {Lu},\ and\ \citenamefont
  {Zhou}}]{Zhang:2026gdk}%
  \BibitemOpen
  \bibfield  {author} {\bibinfo {author} {\bibfnamefont {C.}~\bibnamefont
  {Zhang}}, \bibinfo {author} {\bibfnamefont {Y.}~\bibnamefont {Gong}},
  \bibinfo {author} {\bibfnamefont {X.}~\bibnamefont {Lu}},\ and\ \bibinfo
  {author} {\bibfnamefont {W.}~\bibnamefont {Zhou}},\ }\bibfield  {title}
  {\bibinfo {title} {{The point-particle-limit effective-source approach for
  computing gravitational self-force in the Lorenz gauge}},\ }\href@noop {} {\
  (\bibinfo {year} {2026})},\ \Eprint {https://arxiv.org/abs/2603.27284}
  {arXiv:2603.27284 [gr-qc]} \BibitemShut {NoStop}%
\bibitem [{\citenamefont {Lu}\ \emph {et~al.}(2026)\citenamefont {Lu},
  \citenamefont {Gong}, \citenamefont {Zhang}, \citenamefont {Zhang},\ and\
  \citenamefont {Zhou}}]{Lu:2026jzh}%
  \BibitemOpen
  \bibfield  {author} {\bibinfo {author} {\bibfnamefont {X.}~\bibnamefont
  {Lu}}, \bibinfo {author} {\bibfnamefont {Y.}~\bibnamefont {Gong}}, \bibinfo
  {author} {\bibfnamefont {B.}~\bibnamefont {Zhang}}, \bibinfo {author}
  {\bibfnamefont {C.}~\bibnamefont {Zhang}},\ and\ \bibinfo {author}
  {\bibfnamefont {W.}~\bibnamefont {Zhou}},\ }\bibfield  {title} {\bibinfo
  {title} {{Frequency-domain extended-effective-source gravitational self-force
  for eccentric Schwarzschild orbits}},\ }\href@noop {} {\  (\bibinfo {year}
  {2026})},\ \Eprint {https://arxiv.org/abs/2609.02635} {arXiv:2609.02635
  [gr-qc]} \BibitemShut {NoStop}%
\bibitem [{\citenamefont {Rosa}\ and\ \citenamefont
  {Dolan}(2012)}]{Rosa:2011my}%
  \BibitemOpen
  \bibfield  {author} {\bibinfo {author} {\bibfnamefont {J.~G.}\ \bibnamefont
  {Rosa}}\ and\ \bibinfo {author} {\bibfnamefont {S.~R.}\ \bibnamefont
  {Dolan}},\ }\bibfield  {title} {\bibinfo {title} {{Massive vector fields on
  the Schwarzschild spacetime: quasi-normal modes and bound states}},\ }\href
  {https://doi.org/10.1103/PhysRevD.85.044043} {\bibfield  {journal} {\bibinfo
  {journal} {Phys. Rev. D}\ }\textbf {\bibinfo {volume} {85}},\ \bibinfo
  {pages} {044043} (\bibinfo {year} {2012})},\ \Eprint
  {https://arxiv.org/abs/1110.4494} {arXiv:1110.4494 [hep-th]} \BibitemShut
  {NoStop}%
\bibitem [{\citenamefont {Heffernan}\ \emph {et~al.}(2012)\citenamefont
  {Heffernan}, \citenamefont {Ottewill},\ and\ \citenamefont
  {Wardell}}]{Heffernan:2012su}%
  \BibitemOpen
  \bibfield  {author} {\bibinfo {author} {\bibfnamefont {A.}~\bibnamefont
  {Heffernan}}, \bibinfo {author} {\bibfnamefont {A.}~\bibnamefont
  {Ottewill}},\ and\ \bibinfo {author} {\bibfnamefont {B.}~\bibnamefont
  {Wardell}},\ }\bibfield  {title} {\bibinfo {title} {{High-order expansions of
  the Detweiler-Whiting singular field in Schwarzschild spacetime}},\ }\href
  {https://doi.org/10.1103/PhysRevD.86.104023} {\bibfield  {journal} {\bibinfo
  {journal} {Phys. Rev. D}\ }\textbf {\bibinfo {volume} {86}},\ \bibinfo
  {pages} {104023} (\bibinfo {year} {2012})},\ \Eprint
  {https://arxiv.org/abs/1204.0794} {arXiv:1204.0794 [gr-qc]} \BibitemShut
  {NoStop}%
\bibitem [{\citenamefont {Heffernan}\ \emph {et~al.}(2018)\citenamefont
  {Heffernan}, \citenamefont {Ottewill}, \citenamefont {Warburton},
  \citenamefont {Wardell},\ and\ \citenamefont {Diener}}]{Heffernan:2017cad}%
  \BibitemOpen
  \bibfield  {author} {\bibinfo {author} {\bibfnamefont {A.}~\bibnamefont
  {Heffernan}}, \bibinfo {author} {\bibfnamefont {A.~C.}\ \bibnamefont
  {Ottewill}}, \bibinfo {author} {\bibfnamefont {N.}~\bibnamefont {Warburton}},
  \bibinfo {author} {\bibfnamefont {B.}~\bibnamefont {Wardell}},\ and\ \bibinfo
  {author} {\bibfnamefont {P.}~\bibnamefont {Diener}},\ }\bibfield  {title}
  {\bibinfo {title} {{Accelerated motion and the self-force in Schwarzschild
  spacetime}},\ }\href {https://doi.org/10.1088/1361-6382/aad420} {\bibfield
  {journal} {\bibinfo  {journal} {Class. Quant. Grav.}\ }\textbf {\bibinfo
  {volume} {35}},\ \bibinfo {pages} {194001} (\bibinfo {year} {2018})},\
  \Eprint {https://arxiv.org/abs/1712.01098} {arXiv:1712.01098 [gr-qc]}
  \BibitemShut {NoStop}%
\bibitem [{\citenamefont {Dolan}\ and\ \citenamefont
  {Barack}(2013)}]{Dolan:2012jg}%
  \BibitemOpen
  \bibfield  {author} {\bibinfo {author} {\bibfnamefont {S.~R.}\ \bibnamefont
  {Dolan}}\ and\ \bibinfo {author} {\bibfnamefont {L.}~\bibnamefont {Barack}},\
  }\bibfield  {title} {\bibinfo {title} {{Self-force via $m$-mode
  regularization and 2+1D evolution: III. Gravitational field on Schwarzschild
  spacetime}},\ }\href {https://doi.org/10.1103/PhysRevD.87.084066} {\bibfield
  {journal} {\bibinfo  {journal} {Phys. Rev. D}\ }\textbf {\bibinfo {volume}
  {87}},\ \bibinfo {pages} {084066} (\bibinfo {year} {2013})},\ \Eprint
  {https://arxiv.org/abs/1211.4586} {arXiv:1211.4586 [gr-qc]} \BibitemShut
  {NoStop}%
\end{thebibliography}
\end{document}